\documentclass[10pt]{article}

\usepackage{amsmath}
\usepackage{amssymb}

\usepackage{graphicx}

\usepackage{cite}

\usepackage{color} 

\pdfoutput=1
\usepackage{setspace} 
\usepackage[labelfont=bf,labelsep=period,justification=raggedright]{caption}

\makeatletter
\renewcommand{\@biblabel}[1]{\quad#1.}
\makeatother

\date{}

\newcommand{\dotp}{\! \cdot \!\!}

\begin{document}

\begin{flushleft}
{\Large
\textbf{Coordinated optimization of visual cortical maps\\
(II) Numerical studies}
}
\\
Lars Reichl$^{1,2,3,4,\ast}$, 
Dominik Heide$^{1,5}$, 
Siegrid L\"owel$^{3,6}$,
Justin C. Crowley$^{7}$, 
Matthias Kaschube$^{8}$,
Fred Wolf$^{1,2,3,4\ast}$ 
\\
\bf{1} Max-Planck-Institute for Dynamics and Self-Organization, G\"ottingen, Germany
\\
\bf{2} Bernstein Center for Computational Neuroscience,  G\"ottingen, Germany 
\\
\bf{3} Bernstein Focus Neurotechnology, G\"ottingen, Germany 
\\
\bf{4} Faculty of Physics, Georg-August University, G\"ottingen, Germany
\\
\bf{5} Frankfurt Institute of Advanced Studies, Frankfurt, Germany
\\
\bf{6} School of Biology, Georg-August University, G\"ottingen, Germany
\\
\bf{7} Carnegie Mellon University, Department of Biological Sciences, Pittsburgh PA, USA
\\
\bf{8} Physics Department and Lewis-Sigler Institute, Princeton University, Princeton NJ, USA
\\
$\ast$ E-mail: reichl@nld.ds.mpg.de; fred@nld.ds.mpg.de
\end{flushleft}

\section*{Abstract}
It is an attractive hypothesis that the spatial structure of visual cortical architecture 
can be explained by the coordinated
optimization of multiple visual cortical maps representing orientation preference (OP), ocular
dominance (OD), spatial frequency, or direction preference.
In part (I) of this study we defined a class of analytically tractable
coordinated optimization models and solved representative examples in
which a spatially complex organization of the orientation preference map
is induced by inter-map interactions. We found that attractor solutions
near symmetry breaking threshold predict a highly ordered map layout and
require a substantial OD bias for OP pinwheel stabilization. Here we examine in
numerical simulations whether such models exhibit biologically more
realistic spatially irregular solutions at a finite distance from threshold and
when transients towards attractor states are considered. We also examine
whether model behavior qualitatively changes when the spatial periodicities of
the two maps are detuned and when considering more than 2 feature
dimensions.
We show that, although maps typically undergo substantial rearrangement,   
no other solutions than pinwheel crystals and stripes dominate in the emerging layouts.
Moreover, pinwheel crystallization takes place on a rather short timescale and can also occur for
detuned wavelengths of different maps.
Increasing
the number of feature maps changes conditions for pinwheel stabilization.
In higher dimensional models we find that that 
stabilization by crystallization can be achieved even without an OD bias.
Our numerical results thus support the view that 
neither minimal energy states nor intermediate transient
states of our coordinated optimization models successfully explain the
spatially irregular architecture of the visual cortex. We discuss several
alternative scenarios and additional factors that may improve the agreement 
between model solutions and biological observations.
%
\section*{Author Summary}
Neurons in the visual cortex of carnivores, primates and their close relatives form 
spatial representations or maps of multiple stimulus features.
In part (I) of this study we theoretically predicted maps that are optima of a variety of optimization principles.
When analyzing the joint optimization of two interacting maps we showed that for different 
optimization principles the resulting optima show a stereotyped, spatially perfectly periodic layout.
Experimental maps, however, are much more irregular.
In particular, in case of orientation columns it was found that different species show an apparently species 
invariant statistics of point defects, so-called pinwheels.
In this paper, we numerically investigate whether the spatial features of the stereotyped 
optima described in part (I) are expressed on a biologically relevant timescale 
and whether other, spatially irregular, long-living states emerge that better reproduce the experimentally
observed statistical properties of orientation maps.
Moreover, we explore whether the coordinated 
optimization of more than two maps can lead to spatially irregular optima.
\section*{Introduction}
In the primary visual cortex of primates and carnivores, functional architecture can be characterized
by maps of various stimulus features such as orientation preference (OP), ocular dominance (OD), 
spatial frequency, or direction preference 
\cite{Grinwald,Blasdel3,Swindale3,Grinwald1,Bartfeld,Blasdel4,Blasdel,Grinwald5,Bosking3,Grinwald_Dir,
Chapman,Rao,Bosking,Das2,Loewel,Stryker2,White,Galuske,Casagrande,White_Dir,White3}.
Many attempts have been made to explain and understand the spatial organization of these maps as
optima of specific energy functionals the brain minimizes either during development 
or on evolutionary timescales \cite{Wolf3,Hoffsummer,Goodhill,Goodhill2,Sur5,
Schulten,Obermayer,Sur2,Swindale1,Swindale5,Bednar,Cho1,Cho2,Tanaka,Miller,Pierre,Grossberg}.
In part (I) of this study we presented an analytical approach to study the coordinated 
optimization of interacting pairs of visual cortical maps where 
maps are described by real and complex valued order parameter fields \cite{Reichl5} .
We used symmetry considerations to derive a classification and parametrization of conceivable
inter-map coupling energies and identified a representative set of inter-map coupling terms: 
a gradient-type and a product-type coupling energy which both can enter with different power in the dynamics.
Examining 
this set of inter-map coupling energies was further motivated by the experimentally 
observed geometric 
relationships between cortical maps \cite{Bonhoeffer,Loewel,Blasdel,Bartfeld,Loewel7,Sur5,Casagrande}.
We examined the impact of these coupling energies in a 
system of coupled Swift-Hohenberg equations
which without coupling lead to regular stripe patterns in the complex valued order parameter field.
Our analytical results demonstrated that various types of inter-map coupling energies can induce the formation
of defect structures, so-called pinwheels, in the complex order parameter field describing
the OP map.
For solutions that can become optima of the model, pinwheels are arranged on 
regular periodic lattices such as
rhombic pinwheel crystals (rPWCs) or hexagonal pinwheel crystals (hPWCs).
Our analysis focused on the optimization of a single pair of feature maps where the
complex valued map represents the OP map and the real map represents the OD map.
For this case we presented a complete characterization of the stable 
OP and OD patterns, stripe-like solutions,
rhombic and hexagonal crystal patterns predicted by the coordinated optimization models.
In all analyzed models pinwheel crystallization was conditional on a substantial bias in the response
properties of the co-evolving real-valued map.\\
The pinwheel crystals we obtained, although beautiful and easy to 
characterize, strongly deviate from the spatially irregular design observed for 
OP maps in the visual cortex \cite{Blasdel3,Swindale3,Grinwald1}.
A large scale empirical study of the arrangement of pinwheel positions and their spatial densities 
in the visual cortex of three species widely separated in mammalian
evolution recently showed that orientation maps although spatially
irregular precisely conform with apparently species insensitive
quantitative layout rules \cite{Kaschube6}. In particular, it was found that not only the
mean density of pinwheels but also number fluctuations over a wide range
of spatial scales and local next neighbor arrangements within individual
hypercolumns agree in their statistical features across species with an
accuracy in the range of a few percent \cite{Kaschube6}.
In contrast to the large variability of local map layouts in experimentally 
observed maps 
\cite{Grinwald,Blasdel3,Swindale3,Grinwald1,Bartfeld,Blasdel4,Blasdel,Grinwald5,Bosking3,Grinwald_Dir,
Chapman,Rao,Bosking,Das2,Loewel,Stryker2,White,Galuske,Casagrande,White_Dir,White3}, the
pinwheel crystals found in the coordinated optimization models introduced in part (I)
show a regular and stereotyped structure.
Quantitatively, all PWC solutions that we found exhibit a large pinwheel density 
of about 3.5 or even 5.2 pinwheels 
per hypercolumn. 
In contrast, for experimental OP maps the average pinwheel density was found to be 
statistical indistinguishable
from the mathematical constant $\pi$ up to a precision of 2\% \cite{Kaschube6,Miller4,Stevens}.
Our previous analytical results on optima thus raise the question how coordinated optimization 
of OP and OD maps can be reconciled with the experimentally observed layout rules of orientation maps. \\
From a general perspective, one might suspect that 
the crystalline layout of local minima and optima results from the restrictions of the applied
perturbation method which allowed us to study optima analytically.
Furthermore, results might be expected to change substantially if one would consider 
the coordinated optimization of more than two feature maps.
Examining this aspect is demanded because of the presence of multiple
feature maps in the visual cortex of primates and carnivores and by the
general expectation that geometrical rules coordinating map layout might
be the harder to satisfy the more maps are simultaneously optimized.
Finally, when studying optima of a particular optimization principle we disregarded 
transient states that could in principle dominate developmental optimization 
on biologically relevant timescales and are expected to be more irregular. 
Analytical results were obtained using a perturbative treatment close to the 
pattern forming threshold.
This perturbative treatment, however, gives no information on the speed with which singularities
that were initially generated during spontaneous symmetry breaking will crystallize into highly
ordered arrays.
It is conceivable that this process may occur on very long timescales. If this was the case, 
developmental optimization may lead to long-lived spatially irregular states that are transients
towards regular patterns that would be reached after very long times or potentially never.\\
%
In part (I) we showed that one can neglect the 
backreaction of the OP map onto the OD map if the OD map is 'dominant' i.e. their
amplitudes are much larger than those of the OP map.
This can be achieved for a sufficiently small ratio of their distances to threshold.
This finding raises questions that cannot be addressed purely perturbatively.
Do the observed local minima and optima of the analyzed optimization principles
persist when taking the backreaction into account or when considering map formation
further from the pattern formation threshold?
Besides the influence of the backreaction the full dynamical system receives 
additional corrections. 
There are higher order corrections to the uncoupled amplitude equations which can become important 
for finite bifurcation parameters neglected in \cite{Reichl5}.
In part (I) of this study we also assumed equal periodicities of the two interacting maps.
Experiments suggest that the different periodicities in the layout of OP and OD maps
can have an impact on the map layout \cite{Kaschube1,Kaschube2,Blasdel}.
It is thus also interesting to explore whether and how a detuning of typical periodicities 
affects optimal layouts and whether it can lead to spatially
irregular maps.\\
To assess these issues we generalized the field dynamics to describe the coordinated optimization of 
coupled complex valued and several real valued scalar fields.
From a practical point of view, the analyzed phase diagrams and 
pattern properties indicate that the higher order gradient-type coupling
energy is the simplest and most convenient choice for constructing models that 
reflect the correlations of map layouts in the visual cortex.
For this coupling, intersection angle statistics are reproduced well, pinwheels can
be stabilized, and pattern collapse cannot occur.
In this study we thus numerically analyzed the 
dynamics of coordinated optimization focussing on the high order gradient-type inter-map coupling energy.
We use a fully implicit integrator based on the Crank-Nicolson scheme 
and a Newton-Krylow solver.
In numerical simulations we characterize the kinetics and conditions for pinwheel crystallization and the
creation of pinwheels from a pinwheel-free initial pattern.
The latter is a sufficient but not necessary criterion for systems in which a pinwheel-rich state 
corresponds to an energetically favored state. 
As we point out this criterion can be easily assessed in models of arbitrary complexity 
that otherwise evade analytical treatment.
We further explored the impact of inter-map wavelength differences, as observed in certain species, 
on the structure of the resulting solutions.
Finally, we extended the presented models to explore the coordinated optimization of more
than two feature maps.
To examine whether the observed quantitative properties can be reproduced in models
for the coordinated optimization of maps we calculated various pinwheel statistics during optimization.
We find that spatially irregular patterns decay relatively fast into locally crystalline arrays.
Further long-term rearrangement mainly leads to the emergence of long-range
spatial alignment of local crystalline arrangements.
We showed that the previous finding that OD stripes are not able to stabilize pinwheels
generalizes to the case of detuned wavelengths.
The observation that the
coordinated optimization of two interacting maps leads to spatially 
perfectly periodic optima is also robust to detuned typical wavelengths
and to the inclusion of more than two feature maps.
Our results suggest that the coordinated optimization of multiple maps that would in isolation
exhibit spatially perfectly periodic optimal layouts on its own 
cannot explain the experimentally
observed spatially irregular design of OP maps in the visual cortex and its quantitative aspects.
We consider alternative scenarios and 
propose ways to incorporate inter-map relations and joint optimization
in models in which the optimal OP map layout is intrinsically irregular already 
for vanishing inter-map coupling.
\section*{Results}\label{sec:Results} 
\setcounter{section}{1}
\subsection{A dynamical systems approach}
\begin{figure}[tb]
\includegraphics[width=\linewidth]{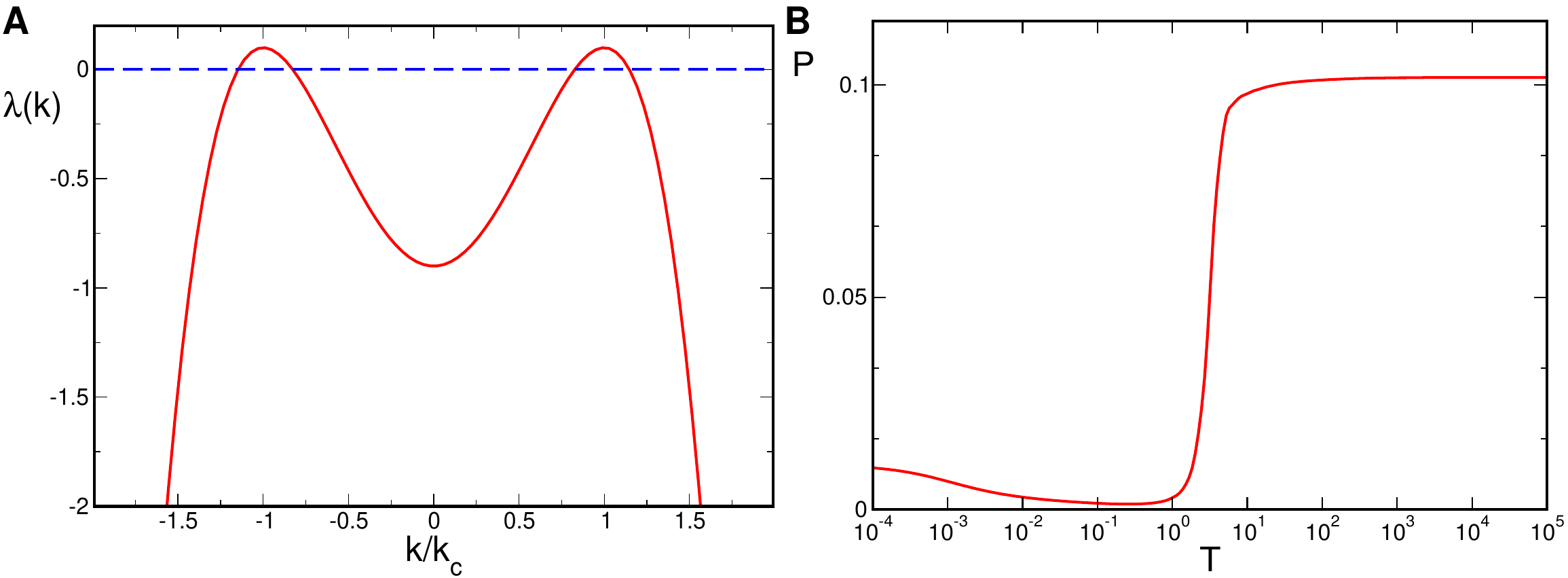} 
\caption{ \textbf{Swift-Hohenberg equation},
\textbf{\sffamily{A}}
Cross section through the spectrum $\lambda(k)$ of the
Swift-Hohenberg operator Eq.~(\ref{eq:LinearSHkspace}), $r=0.1$.
\textbf{\sffamily{B}} Time evolution of the Power, Eq.~(\ref{eq:Power}), for spatial white noise
initial conditions.
}
\label{fig:SwiftHohenberg}
\end{figure}

We model the response properties of neuronal populations in the visual cortex by two-dimensional
scalar order parameter fields which are either complex valued 
or real valued \cite{Swindale2,Swindale7,Swindale6}.
We consider inter-map coupling between a complex valued map $z(\mathbf{x})$ and one or several real 
valued maps $o_i(\mathbf{x})$.
The complex valued field $z(\mathbf{x})$ can for instance describe OP or
direction preference of a neuron located at position $\mathbf{x}$.
A real valued field $o(\mathbf{x})$ can describe for instance OD or the spatial frequency preference. 
Although we consider a model for the coordinated optimization of general real and complex valued 
order parameter fields we view $z(\mathbf{x})$ as the field of OP throughout this article to aid 
comparison to the biologically observed patterns.
In this case, the pattern of preferred stimulus orientation $\vartheta$ is obtained by
\begin{equation} 
\vartheta(\mathbf{x})=\frac{1}{2} \arg (z).
\end{equation} 
The modulus $|z(\mathbf{x})|$ is a measure of orientation selectivity at cortical location $\mathbf{x}$.\\
OP maps are characterized by so-called \textit{pinwheels}, regions
in which columns preferring all possible orientations are organized around a 
common center in a radial fashion \cite{Swindale2,Ohki,Blasdel5,Bonhoeffer2}.
The centers of pinwheels are point discontinuities of the field $\vartheta(\mathbf{x})$
where the mean orientation preference of nearby columns changes by 90 degrees.
Pinwheels can be characterized by a topological charge $q$ which indicates in particular
whether the orientation preference increases clockwise or counterclockwise around
the pinwheel center,
\begin{equation}
q_i=\frac{1}{2\pi}\oint_{C_i} \nabla \vartheta(\mathbf{x}) d\mathbf{s}\, ,
\end{equation}
where $C_i$ is a closed curve around a single pinwheel center at $\mathbf{x}_i$. 
Since $\vartheta$ is a cyclic variable in the interval $[0,\pi]$ and up to isolated points
is a continuous function of $\mathbf{x}$, $q_i$ can only have values
\begin{equation}
q_i=\frac{n}{2} \, ,
\end{equation}
where $n$ is an integer number \cite{Mermin}. If its absolute value $|q_i|=1/2$, each orientation is 
represented only
once in the vicinity of a pinwheel center.
In experiments, only pinwheels with a topological charge of $\pm 1/2$ have been observed, which are 
simple zeros of the field $z(\mathbf{x})$.\\
In case of a single real valued map $o(\mathbf{x})$ the field can be considered as the field of OD, 
where $o(\mathbf{x})<0$ indicates ipsilateral eye dominance and $o(\mathbf{x})>0$ 
contralateral eye dominance of the neuron located at position $\mathbf{x}$. The magnitude
indicates the strength of the eye dominance and thus the zeros of the 
field corresponding to the borders of OD.\\
As visual cortical maps are described by optima of an energy functional $E$, a formal 
time evolution of these maps that represents the the gradient descent of this energy functional
can be used to obtain predicted map layouts.
The field dynamics thus takes the form
\begin{eqnarray}\label{eq:coupledGeneral}
\partial_t \, z(\mathbf{x},t) &=& F[z(\mathbf{x},t),o_1(\mathbf{x},t),o_2(\mathbf{x},t), \dots] \nonumber \\
\partial_t \, o_i(\mathbf{x},t) &=& G_i[z(\mathbf{x},t),o_1(\mathbf{x},t),o_2(\mathbf{x},t),\dots], 
\end{eqnarray}
where $F[z,o_i]$ and $G_i[z,o_i]$ are nonlinear operators given by
$F[z,o_i]=-\frac{\delta E}{\delta \overline{z}}$, $G_i[z,o_i]=-\frac{\delta E}{\delta o_i}$.
The system then relaxes towards the minima of the energy $E$.
The convergence of this dynamics towards an attractor is assumed to represent
the process of maturation and optimization of the cortical circuitry.
Various biologically detailed models can be cast into the form 
of Eq.~(\ref{eq:coupledGeneral}) \cite{Swindale6,Wolf3,Scherf}.\\
To dissect the impact of inter-map coupling interactions we split the energy functional $E$ into
single field and interaction components $E=E_z+E_{o_i}+\sum_iU_{zo}(z,o_i)+\sum_{i\neq j}
U_{oo}(o_i,o_j)$.
All visual cortical maps are arranged in roughly repetitive patterns of a typical wavelength $\Lambda$ that
may be different for different maps.
We chose $E_z$ to obtain, in the absence of coupling, a well studied model reproducing 
the emergence of a typical wavelength by a pattern forming 
instability, the Swift-Hohenberg model \cite{SwiftHohenberg,Cross}.
This model has been characterized comprehensively in the pattern formation literature and mimics the
behavior of for instance the continuous Elastic Network or the Kohonen model for
orientation selectivity (see \cite{Wolf3}).
We note that many other pattern forming systems occurring in different physical, chemical, and 
biological contexts (see for instance \cite{Busse,Bodenschatz,Soward,Vinals}) 
have been cast into a dynamics of the same form.
Its dynamics in case of the OP map is of the form
\begin{equation}\label{eq:DynamicsSH}
\partial_t \, z(\mathbf{x},t) = \hat{L} z(\mathbf{x},t) -|z|^2z\, , 
\end{equation}
with the linear Swift-Hohenberg operator
\begin{equation}\label{eq:LinearSH}
\hat{L}=r-\left(k_c^2+\Delta \right)^2\, , 
\end{equation}
$k_c=2\pi/\Lambda$, and $\Delta$ the Laplace operator.
In Fourier representation, $\hat{L}$ is diagonal with the spectrum
\begin{equation}\label{eq:LinearSHkspace}
\lambda(k)=r-\left(k_c^2-k^2 \right)^2\, . 
\end{equation}
The spectrum exhibits a maximum at $k=k_c$, see Fig.~\ref{fig:SwiftHohenberg}(a). 
For $r<0$, all modes are damped since $\lambda(k)<0 , \, \forall k$ and only the 
homogeneous state $z(\mathbf{x})=0$ is stable.
This is no longer the case for $r>0$ when modes on the \textit{critical circle} $k=k_c$
acquire a positive growth rate and grow, resulting in patterns with a typical 
wavelength $\Lambda=2\pi/k_c$.
This model exhibits a supercritical bifurcation where the homogeneous state looses 
its stability and spatial pattern emerge.\\
While the linear part of the dynamics establishes a typical wavelength, the nonlinear term
in the dynamics leads to the selection of the final pattern.
In this study we measure time $T=r t$ in units of the intrinsic timescale which is associated with the
growth rate of the linear part Eq.~(\ref{eq:LinearSH}).
Considering the time evolution of Eq.~(\ref{eq:DynamicsSH}) 
initialized with a random OP map and low selectivity (small $|z|$) 
several different stages of the dynamics can be distinguished.
The linear part forces modes on the critical circle to grow with rate $r$ while strongly
suppressing modes off the critical circle when starting from small amplitude 
white noise initial conditions, see Fig.~\ref{fig:SwiftHohenberg}\textbf{\sffamily{A}}.
The OP map becomes more ordered in this linear phase as one dominant wavelength emerges.
The total power of the field is given by
\begin{equation}\label{eq:Power}
P(t)=\langle |z(\mathbf{x},t)|^2 \rangle _\mathbf{x}\, ,
\end{equation}
where $\langle \rangle _{\mathbf{x}}$ denotes spatial average.
The time dependence of the power reflects the different growth rates among modes.
The time evolution of the power is depicted in Fig.~\ref{fig:SwiftHohenberg}\textbf{\sffamily{B}}.
Initially, the power decreases slightly due to the suppression 
of modes outside the circle of positive growth rate.
At $T \approx 1$ there is a rapid increase and then a saturation of the power.  
The amplitudes of the Fourier modes reach their stationary values and $P\propto r$.
At this stage of the evolution the influence of the nonlinear part becomes comparable to that of the linear part.
Once the modes saturate the phase of nonlinear competition between the active modes along with 
a reorganization of the structure of the OP map starts.
The competition between active modes leads to pattern selection.
The final pattern then consists of distinct modes in Fourier space \cite{Manneville,Cross}. 
Once the active modes are selected a relaxation of their phases takes place.\\
Inter-map coupling can influence the time evolution on all stages of the development
depending on whether this coupling affects only the nonlinear part or also the linear one. 
When incorporating additional maps into the system in all cases we rescaled the
dynamics by the bifurcation parameter of the OP map i.e. $T=r_z t$.
The coupled dynamics we considered is of the form
\begin{eqnarray}\label{eq:DynamicsSHcoupled_ref}
\partial_t \, z(\mathbf{x},t) &=& \hat{L}_z \, z(\mathbf{x},t) -|z|^2z 
-\frac{\delta U}{\delta \overline{z}} \nonumber \\ 
\partial_t \, o_i(\mathbf{x},t) &=& \hat{L}_{o_i} \, o_i(\mathbf{x},t) 
-o_i^3 -\frac{\delta U}{\delta o_i}+\gamma \, ,
\end{eqnarray}
where $\hat{L}_{\{o,z\}}=r_{\{o_i,z\}} -\left(k_{c,\{o_i,z\}}^2+\Delta \right)^2$, and $\gamma$ is 
a constant.
To account for the differences in the dominant wavelengths of the patterns we chose two
typical wavelengths $\Lambda_z=2\pi/k_{c,z}$ and $\Lambda_{o_i}=2\pi/k_{c,o_i}$.
Since in cat visual cortex the typical wavelength for OD and OP maps are approximately the 
same \cite{Kaschube1,Kaschube2} i.e. $k_{c,o}=k_{c,z}=k_c$
the Fourier components of the emerging pattern in this case are located on a common critical 
circle, see however \cite{Diao}.
The dynamics of $z(\mathbf{x},t)$ and $o_i(\mathbf{x},t)$ are coupled by interaction terms which
can be derived from a coupling energy $U$.
Many optimization models of the form presented in Eq.~(\ref{eq:coupledGeneral}) have been 
studied \cite{Wolf3,Hoffsummer,Goodhill,Goodhill2,Sur5,
Schulten,Obermayer,Sur2,Swindale1,Swindale5,Bednar,Cho1,Cho2,Tanaka,Miller,Pierre,Grossberg}.
The concrete dynamics in Eq.~(\ref{eq:DynamicsSHcoupled_ref}) is the simplest which
in the uncoupled case leads to pinwheel-free OP stripe patterns and to a stripe-like or patchy layout of
the co-evolving real valued fields.\\
As revealed by the symmetry-based classification of coupling energies 
\begin{equation}\label{eq:Energy_all}
U=\alpha \, o^2|z|^2+\beta \, |\nabla z \dotp \nabla o|^2 +\tau \, o^4 |z|^4 +\epsilon\, |\nabla z \dotp \nabla o|^4 \, ,
\end{equation}
parametrizes a representative family of biologically plausible coupling energies for a single 
real valued map $o$, see part (I), \cite{Reichl5}.\\
The numerical integration scheme to solve Eq.~(\ref{eq:DynamicsSHcoupled_ref}) is detailed 
in the Methods part.
Initial conditions for the OD map were chosen as stripe, hexagonal, or constant patterns 
plus Gaussian white noise.
Initial conditions for the OP map are either pinwheel-free OP stripes
or band-pass filtered Gaussian white noise for which the average pinwheel density is
bounded from below by the constant $\pi$ \cite{Wolf3}.
For numerical analysis we focused on the high order gradient-type inter-map
coupling energy.
This energy can reproduce all qualitative 
relationships found between OP and OD maps, does
not suffer from potential OP map suppression, and leads to a relatively simple phase 
diagram for two interacting maps near threshold.\\
\subsection{Final states}
\begin{figure}
\includegraphics[width=\linewidth]{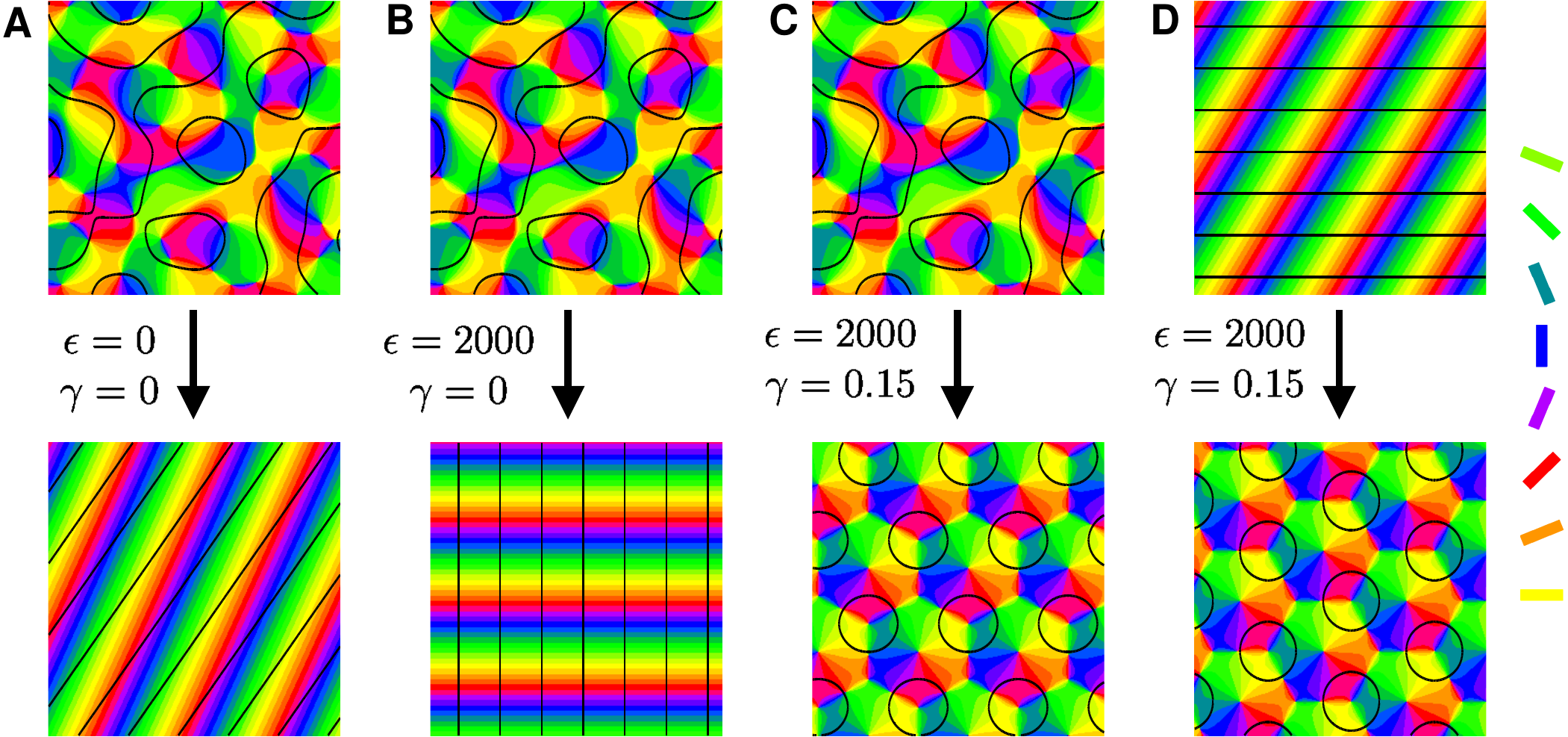}
\caption{
\textbf{Pinwheel annihilation, preservation, and generation} in numerical simulations for different 
strengths of inter-map coupling $\epsilon$ and OD bias $\gamma$.
Color code of OP map with zero contours of OD map superimposed. 
\textbf{\sffamily{A}} $\gamma=0,\epsilon=0$
\textbf{\sffamily{B}} $\gamma=0,\epsilon=2000$
\textbf{\sffamily{C}} and \textbf{\sffamily{D}} $\gamma=0.15,\epsilon=2000$.
Initial conditions identical in \textbf{\sffamily{A}} - \textbf{\sffamily{C}}, 
$r_o=0.2, r_z=0.02, T_{f}=10^4, k_{c,o}=k_{c,z}$.}
\label{fig:Numerics}
\end{figure}

In part (I)  we calculated phase diagrams for different inter-map coupling energies \cite{Reichl5}. 
In all cases, hexagonal PWCs can be stabilized only in case of OD hexagons.
We tested these results numerically. 
Numerical simulations of the dynamics Eq.~(\ref{eq:DynamicsSHcoupled_ref}) with the coupling 
energy 
\begin{equation}\label{eq:Energy_used}
U=\epsilon\, |\nabla z \dotp \nabla o|^4 \, ,
\end{equation}
are shown in Fig.~\ref{fig:Numerics}.
All remaining inter-map coupling energies in Eq.~(\ref{eq:Energy_all}) are assumed to be zero.
The initial conditions and final states are shown for different bias terms $\gamma$ 
and inter-map coupling strengths $\epsilon$. 
Note the logarithmic time scales. Pinwheel densities rapidly diverge from values near 3.1 as soon as the map
exhibits substantial power $1<T<10$.
We observed that for a substantial contralateral bias and above a
critical inter-map coupling pinwheels are preserved for all times or
are generated if the initial condition is pinwheel-free.
Without a contralateral bias the final states were pinwheel-free stripe solutions 
irrespective of the strength of the inter-map coupling.
\subsection{Kinetics of pinwheel crystallization}\label{sec:PWkinematics}
\begin{figure}[tb]
\centering
\includegraphics[width=\linewidth]{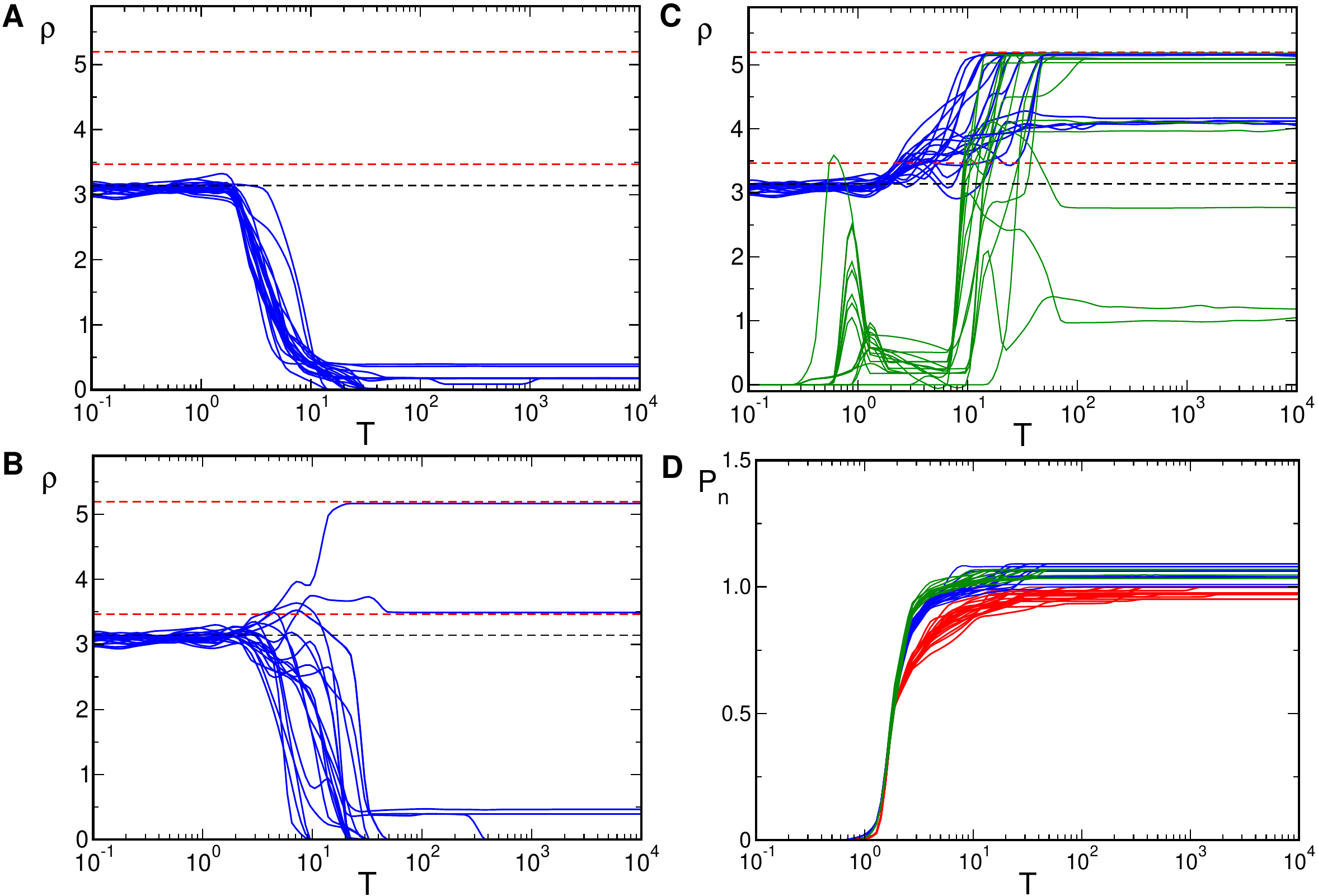}
\caption{\textbf{Time evolution of the pinwheel density, $U=\epsilon\, |\nabla z \dotp \nabla o|^4 $}, 
$r_z=0.05, r_o=0.25, \gamma=0.15$.
For each parameter set \textbf{\sffamily{A}}-\textbf{\sffamily{C}} simulations in blue started from an identical set 
of $20$ initial conditions.
Red dashed lines: $\rho=4\cos(\pi/6)$ and
$\rho=6\cos(\pi/6)$, black dashed line: $\rho=\pi$.  
\textbf{\sffamily{A}} $\epsilon=0$ \textbf{\sffamily{B}} 
$\epsilon=200$ \textbf{\sffamily{C}} $\epsilon=2000$.
\textbf{\sffamily{D}} Normalized power of OP map, $\epsilon=0$ (red), $\epsilon=200$ (blue), and $\epsilon=2000$ (green).
In green \textbf{\sffamily{C}}: OD and OP stripes as initial conditions. 
Parameters: $128 \times 128$ mesh, $\Gamma=22$.
}
\label{fig:PWkin} 
\end{figure}

To characterize the process of pinwheel annihilation, preservation, and creation during progressive
map optimization we calculated the pinwheel density as well as various other 
pinwheel statistics (see Methods) during the convergence of patterns to attractor states.
The time evolution of the pinwheel density is shown in Fig.~\ref{fig:PWkin}.
In the uncoupled case ($\epsilon=0$) most of the patterns decayed into a stripe solution
and their pinwheel density dropped to a value near zero. At small coupling strengths ($\epsilon=200$) 
the pinwheel density
converged either to zero (stripes), to values near 3.5 for the rPWC (see part (I), Figure S6), or
to approximately 5.2 for the contra-center hPWC (see part (I), Figure S7). At high inter-map coupling ($\epsilon=2000$) 
pinwheel free stripe patterns formed neither from pinwheel rich nor from pinwheel free initial conditions.
In this regime the dominant layout was the contra-center hPWC. When starting from OD and OP stripes, 
see Fig.~\ref{fig:PWkin}\textbf{\sffamily{C}} (green lines), the random orientation between the stripes first evolved towards a 
perpendicular orientation ($T\approx 1$).
This lead to a transient increase in the pinwheel density. At the time ($T\approx 10$) where the OD stripes 
dissolve towards OD hexagons hPWC solutions formed and the pinwheel density reached 
its final value.\\ 
Regions of hPWC layout
can however be inter-digitated with long lived rPWC solutions and stripe domains.
Figure~\ref{fig:PWkin}\textbf{\sffamily{D}} shows the time course of the normalized power 
$P_n(t)=\langle |z(\mathbf{x},t)_{\operatorname{dyn}}|^2 \rangle _\mathbf{x} / \langle 
|z(\mathbf{x},t)_{\operatorname{th}}|^2 
\rangle _\mathbf{x}$, where $\langle \rangle _\mathbf{x}$ denotes spatial average. 
The field $z_{\operatorname{th}}$ is obtained from the solution of the amplitude equations (see \cite{Reichl5}) 
while $z_{\operatorname{dyn}}$ is the field obtained from the simulations.
Starting from a small but nonzero power the amplitudes grew and saturated after $T \approx 1 $. 
When the amplitudes were saturated the selection of the final pattern started.
Quantitatively, we found that with weak backreaction the critical coupling strengths were slightly
increased compared to their values in the limit $r_z \ll r_o$. 
Snapshots of the simulation leading to the hPWC solutions at three 
time frames are shown in Fig.~\ref{fig:Snapshot.pdf}.
\begin{figure}[tb]
\centering
\includegraphics[width=.7\linewidth]{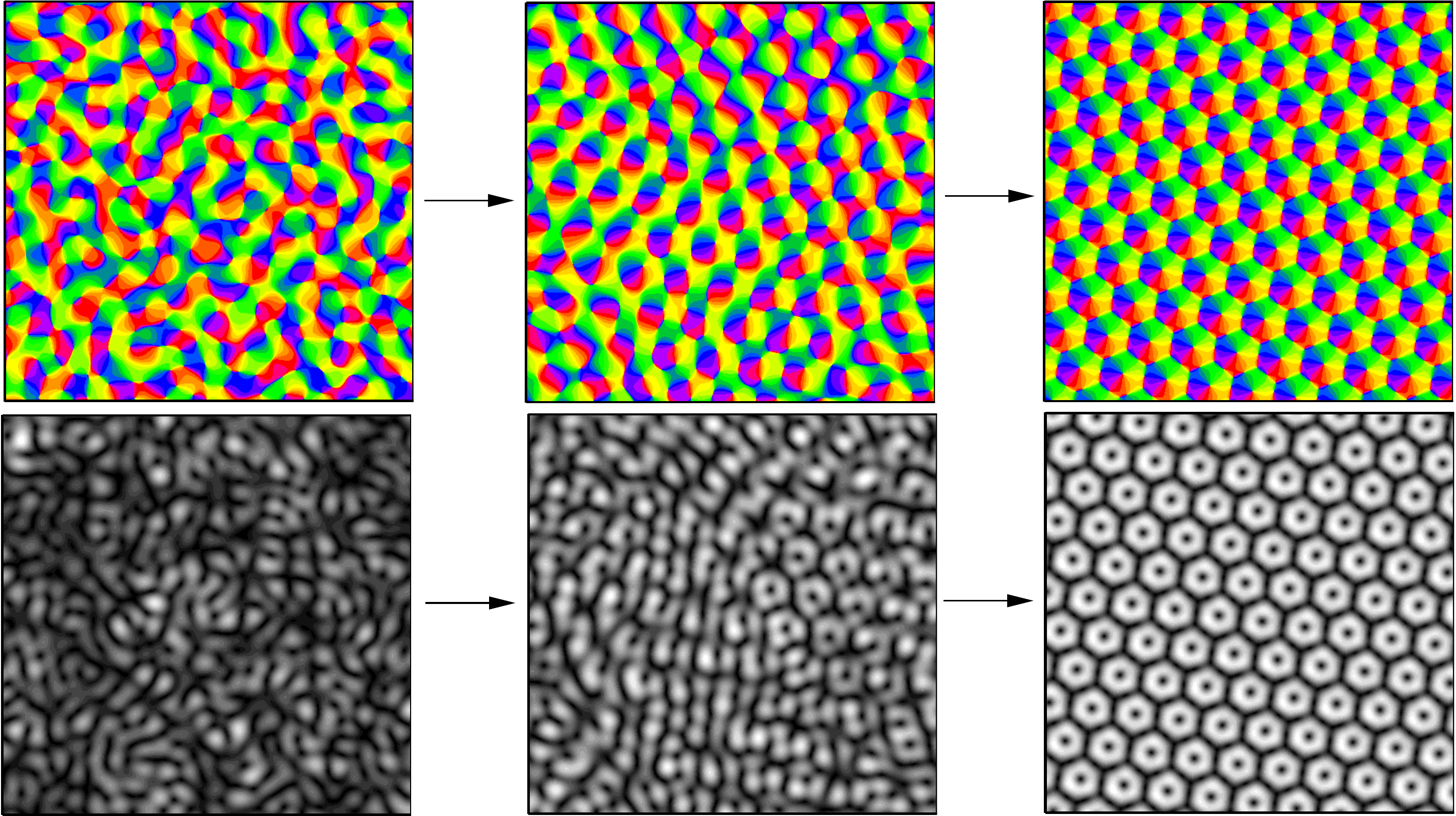}
\caption{\textbf{Snapshots of the pinwheel crystallization process.} Top panel: OP map, bottom 
panel: selectivity $|z(\mathbf{x})|$.
Left: $T=0.01$, middle: $T=0.8$, right: $T=T_f=10^4$. 
Parameters as in Fig.~\ref{fig:PWkin}(c), $\epsilon=2000, \gamma=0.15$}
\label{fig:Snapshot}
\end{figure}
Already at $T\approx0.8$ a substantial  rearrangement of the pattern took place and 
one can identify different domains in the pattern that are locally highly stereotyped.\\
For the time evolution of the maps we also calculated the distributions of pinwheel next-neighbor distances $d$,
measured in units of the column spacing $\Lambda$. The distributions of distances for  
simulations leading to rhombic and hPWC solutions are shown in Fig.~\ref{fig:PWdistance}.
\begin{figure}[tb]
\begin{center}
\includegraphics[width=0.9\linewidth]{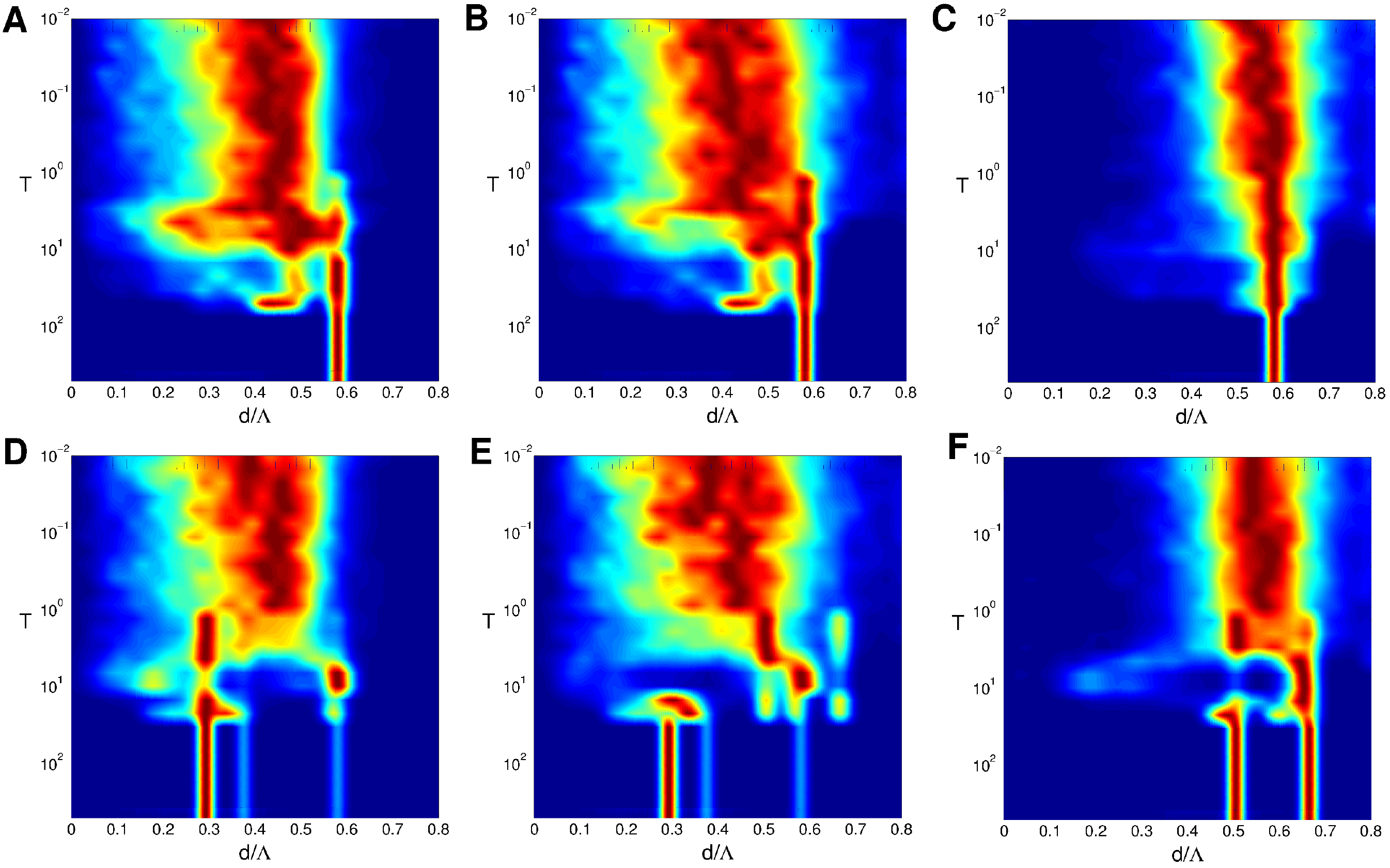}
\end{center}
\caption{\textbf{Distribution of nearest neighbor pinwheel distances during development.} 
\textbf{\sffamily{A-C}} rPWC 
\textbf{\sffamily{D-F}} hPWC.
Distance to the next pinwheel of arbitrary \textbf{\sffamily{A,D}}, opposite \textbf{\sffamily{B,E}}, and
equal \textbf{\sffamily{C,F}} topological charge.
Parameters as in Fig.~\ref{fig:PWkin}\textbf{\sffamily{B}}. 
}
\label{fig:PWdistance}
\end{figure}
They are characterized by three stages in the evolution of the pinwheel distances.
At early stages of the evolution ($10^{-2}\leq T$) there is a continuous distribution 
starting approximately linearly from $d=0$.
At the time where the amplitudes saturated ($T\approx 1$) the distribution of pinwheel distances became
very inhomogeneous. Different domains with stripe-like, rhombic, or hexagonal patterns appeared
(see also Fig.~\ref{fig:PWCreatAnnihi}\textbf{\sffamily{C,D}}) until for $T>10$ the 
rhombic or hexagonal pattern took over 
the entire pattern.\\
As pinwheels carry a topological charge we could divide the distributions according to distances
between pinwheels of the same charge or according to distances between pinwheels of
the opposite charge.
In Fig.~\ref{fig:PWdistance_last} we present pinwheel distances for the final 
states of the dynamics.
\begin{figure}
\begin{center}
\includegraphics[width=.7\linewidth]{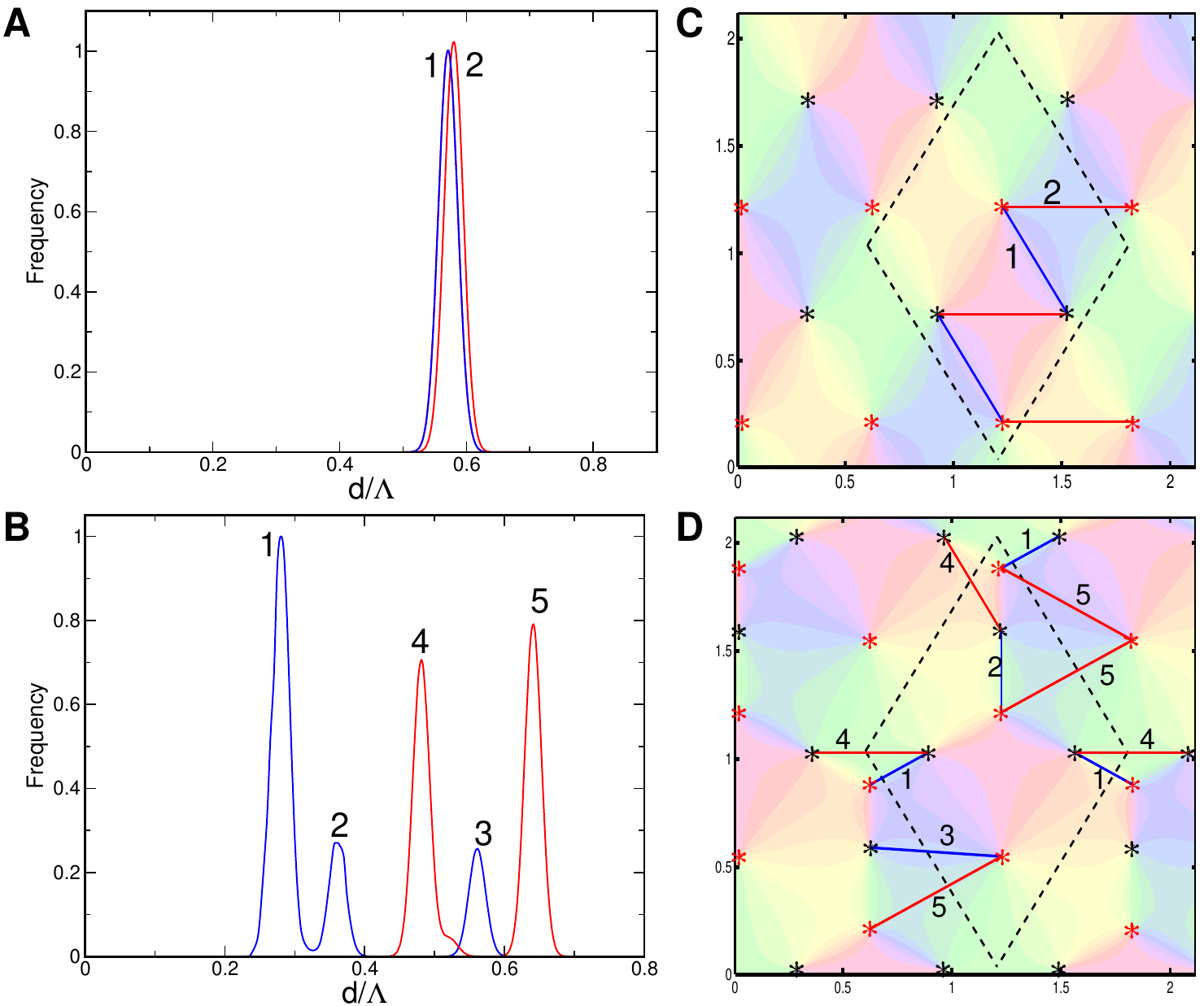}
\end{center}
\caption{\textbf{Distribution of nearest neighbor distances for final states} ($T=T_f=10^4$). 
\textbf{\sffamily{A}} rPWC, \textbf{\sffamily{B}} hPWC 
with pinwheels of equal (red) and opposite (blue) charge.
\textbf{\sffamily{C}} and \textbf{\sffamily{D}} Illustration of occurring pinwheel distances.
Pinwheels are marked with star symbols according to their charge. Units are given in $\Lambda$.
Parameters as in Fig.~\ref{fig:PWkin}\textbf{\sffamily{B}}. 
}
\label{fig:PWdistance_last}
\end{figure}
In case of the rhombic solutions there is only a single pinwheel to pinwheel distance with 
$d=1/\sqrt{3} \Lambda \approx 0.58\, \Lambda$. 
In numerical simulations small variations in the amplitudes lead to a slightly larger distance between
pinwheels of equal charge than between pinwheels of opposite charge. Therefore their distance 
distributions do not collapse exactly, see Fig.~\ref{fig:PWdistance_last}\textbf{\sffamily{A}}.
In case of the hPWC there are three peaks at 
$d\approx 0.28\, \Lambda, d\approx 0.36\, \Lambda$ and $d\approx 0.56\, \Lambda$ in the 
pinwheel distance distribution of arbitrary charge, see Fig.~\ref{fig:PWdistance_last}\textbf{\sffamily{B}}.
These three peaks all result from distances between pinwheels carrying the opposite charge
while the distance between pinwheels of the same charge shows two peaks at $d\approx 0.48\, \Lambda$
and $d\approx 0.64\, \Lambda$ in the distribution.
The origin of the peaks is indicated in Fig.~\ref{fig:PWdistance_last}\textbf{\sffamily{C}} and 
Fig.~\ref{fig:PWdistance_last}\textbf{\sffamily{D}}.\\
These results confirm that inter-map coupling can induce the stabilization of pinwheels in the OP pattern.
This however does not mean that the pinwheels initially generated by spontaneous symmetry
breaking will be preserved during convergence of the map.
To what extent are the pinwheels in the crystalline OP maps preserved from pinwheels
of the initial OP pattern? To answer this question we calculated the pinwheel annihilation $a(t)$ and
creation $c(t)$ rate during time evolution, see Methods. 
The time evolution of these rates, averaged over 20 simulations leading to 
a hPWC, is shown in Fig.~\ref{fig:PWCreatAnnihi}\textbf{\sffamily{A}}.
\begin{figure}[tb]
\centering
\includegraphics[width=\linewidth]{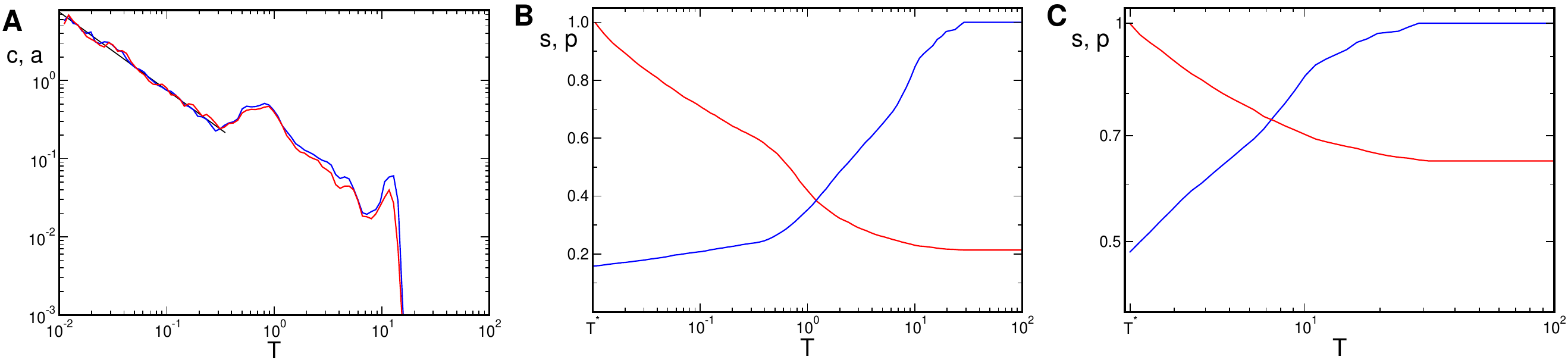}
\caption{\textbf{Pinwheel annihilation and creation.} \textbf{\sffamily{A}} Creation (blue) and annihilation 
(red) rates during time evolution. Fit: $c,a = 0.08/T$ (black line).
\textbf{\sffamily{B,C}} Survival fraction (red) and fraction of preserved pinwheels (blue) compared
to the initial time $T^*=0.01$ \textbf{\sffamily{B}} and $T^*=2$ \textbf{\sffamily{C}}.
Parameters as in Fig.~\ref{fig:PWkin}\textbf{\sffamily{C}}.
}
\label{fig:PWCreatAnnihi}
\end{figure}
We observe that both rates were fairly similar throughout development, with a slightly higher
creation rate in the later stage of development.
During the initial stages of time evolution creation and annihilation rates
decay algebraically $c,a \propto 1/T$. At $T\approx 3$ both rates deviate from this algebraic decay.
From thereon annihilation and creation rates increase, reflecting the nonlinear rearrangement of the pattern.
After $T\approx 15$ no pinwheels are created or annihilated anymore and the pinwheels of the 
final pattern are present.\\
Pinwheels are created and annihilated until a first crystal-like pattern is formed.
How many pinwheels of the initial pattern are still present in the final pattern?
For a given set of pinwheels at an initial time $T=T^*$ we further calculate the fraction $s(t)$
of those pinwheels surviving until time $T$. The fraction of pinwheels present at time $T^*$ that 
survive up to the final time $T=T_{f}$ is given by $p(t)$. 
Both fractions are shown in Fig.~\ref{fig:PWCreatAnnihi}\textbf{\sffamily{B}} for $T^*=0.01$
and in Fig.~\ref{fig:PWCreatAnnihi}\textbf{\sffamily{C}} for $T^*=2$, a time where the power $P(t)$ has almost 
saturated, see Fig.~\ref{fig:PWkin}\textbf{\sffamily{D}}. 
We observed that about 20\% of the initial pinwheels are preserved until the final time and 
therefore most of the pinwheels of the crystal pattern are created during development.
From those pinwheels which are present when the power saturates about 65\% are also present 
in the final pattern.
\subsection{Detuning OD and OP wavelengths: OD stripes}\label{sec:Macaque_part3}
\begin{figure}[bt]
\begin{center}
\includegraphics[width=\linewidth]{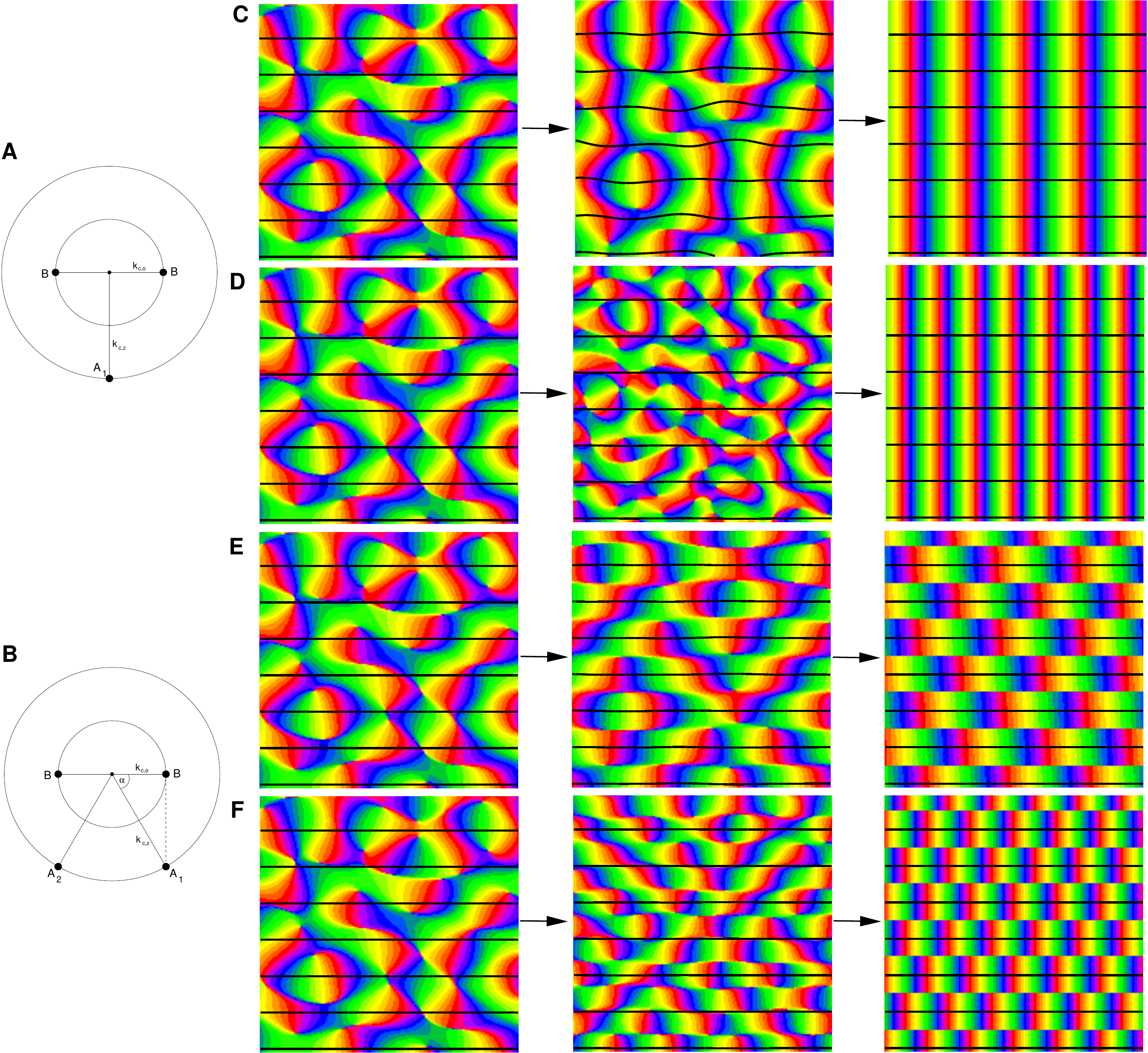} 
\end{center}
\caption{\textbf{Map interactions with detuned wavelengths and OD stripes.}
OD stripes interacting with OP columns where $\Lambda_o/\Lambda_z>1$.
\textbf{\sffamily{A,B}} Illustration of active modes in Fourier space 
with $k_{c,o}<k_{c,z}$, $\alpha=\arccos k_{c,o}/k_{c,z}$.
\textbf{\sffamily{C,D}} $U=\epsilon |\nabla z \dotp \nabla o|^4$, $\epsilon=2000$,
\textbf{\sffamily{E,F}} $U=\tau o^4 |z|^4$, $\tau=2000$,
\textbf{\sffamily{C,E}} $\Lambda_{o}/\Lambda_{z}=1.3$,
\textbf{\sffamily{D,F}} $\Lambda_{o}/\Lambda_{z}=2$.
Left: initial condition, middle: $T=1$, right: $T=5 \cdot 10^4$.
Parameters: $r_z=0.05, r_o=0.2, \Gamma_z=20, 256\times 256$ mesh. 
Initial condition identical in all simulations.
}
\label{fig:OffCircle_part3}
\end{figure}

The analytical results obtained in \cite{Reichl5} as well 
as the previous numerical results (see Fig.~\ref{fig:Numerics}\textbf{\sffamily{B}}) predict 
that OD stripes do not lead to spatially complex patterns and are not capable of stabilizing pinwheels.
In case of gradient-type inter-map couplings the OP map consists of stripes which run
perpendicular to the OD stripes.
In case of the product-type inter-map coupling
high gradient regions of both maps avoid each other by producing again OP stripes
but now running in the same direction as the OD 
stripes. 
In numerical simulations we also investigated the case of OD stripes of larger wavelength 
than OP columns, as is the case in macaque monkey primary visual cortex \cite{Blasdel,Blasdel6}.
In case of a gradient-type inter-map coupling we find that the OD bands are perpendicular 
to the OP bands independent of the 
ratio $\Lambda_{o}/\Lambda_{z}>1$, see Fig.~\ref{fig:OffCircle_part3}\textbf{\sffamily{C,D}}.
In case of the product-type inter-map coupling, if the ratio $\Lambda_{o}/\Lambda_{z}>1$, orientation 
representation does not collapse as it would be the 
case for $\Lambda_{o}=\Lambda_{z}$, see\cite{Reichl5}.
The system, however, again finds a way to put zero contours (Re $z=0$ and Im $z=0$)
along the OD maximum which now is a 
fracture line, see Fig.~\ref{fig:OffCircle_part3}\textbf{\sffamily{E,F}}.
The angle between the active OP and OD modes is given by $\alpha=\arccos k_{c,o}/k_{c,z}$ 
corresponding to the resonance condition 
$\vec{k}_{1,z}-\vec{k}_{2,z}-2\vec{k}_{1,o}=0$, see Fig.~\ref{fig:OffCircle_part3}\textbf{\sffamily{A}}.
\subsection{Detuning OD and OP wavelengths: OD hexagons}\label{sec:Macaque_part3}
\begin{figure}[bt]
\begin{center}
\includegraphics[width=.8\linewidth]{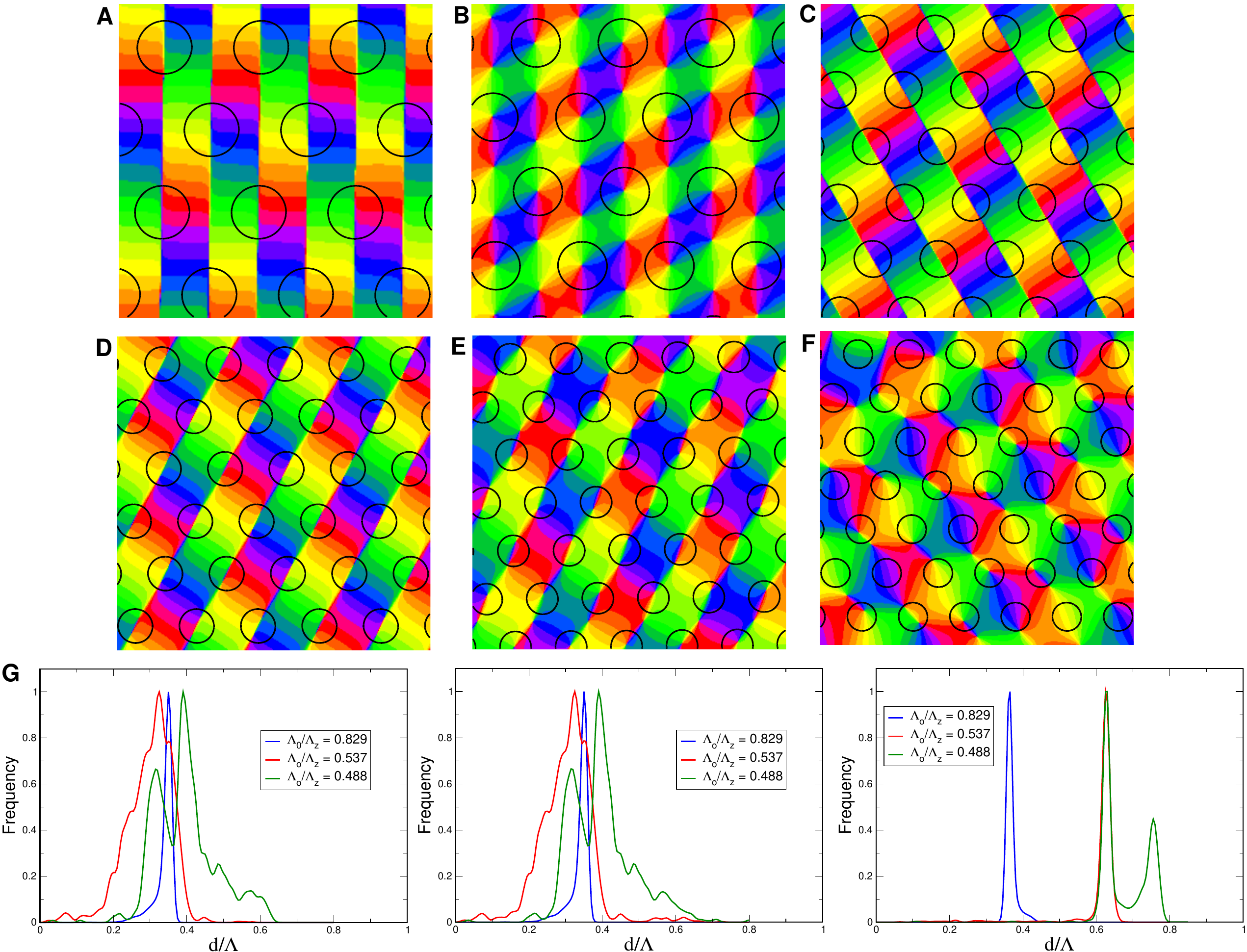} 
\end{center}
\caption{\textbf{Map interactions with detuned wavelength and OD hexagons.}
$U=\epsilon \, |\nabla z  \dotp \nabla o|^4$.
\textbf{\sffamily{A}} $\Lambda_{o}/\Lambda_{z}= 38/41\approx 0.927$,
\textbf{\sffamily{B}} $\Lambda_{o}/\Lambda_{z}= 34/41\approx 0.829$,
\textbf{\sffamily{C}} $\Lambda_{o}/\Lambda_{z}= 26/41\approx 0.634$,
\textbf{\sffamily{D}} $\Lambda_{o}/\Lambda_{z}= 24/41\approx 0.586$,
\textbf{\sffamily{E}} $\Lambda_{o}/\Lambda_{z}= 22/41\approx 0.537$,
\textbf{\sffamily{F}} $\Lambda_{o}/\Lambda_{z}= 20/41\approx 0.488$.
\textbf{\sffamily{G}} Distribution of nearest neighbor pinwheel 
distances of arbitrary (Left), opposite (Middle), and
equal (Right) topological charge.
Parameters: $\gamma=0.15, r_z=0.05, r_o=0.2,\epsilon=2000,
\Gamma_o=41, T_f=10^4$, $256\times 256$ mesh. Initial condition identical in all simulations.
}
\label{fig:PWstab_inc}
\end{figure}

In case of identical wavelengths $k_{c,o}=k_{c,z}$ strong interaction with a system of 
hexagonal OD patches leads to hPWC solutions.
For these solutions pinwheel positions are correlated with OD extrema.
For instance in case of the higher order gradient-type inter-map coupling energy,
for which the contra-center PWC corresponds to the energetic ground state, half of the pinwheels
are located at OD extrema while the remaining half is located near 
OD borders, see Figure S7 of part (I) \cite{Reichl5}. 
If, however, the typical wavelengths of OD and OP patterns are not identical such a precise relationship
cannot be fulfilled in general.
We therefore studied whether a detuning of typical wavelengths can lead to spatially irregular and 
pinwheel rich OP patterns.
In numerical simulations which lead to OD hexagons with a fixed wavelength we varied
the OP wavelength using identical initial conditions.
Wavelength ratios are chosen such that each pattern exhibited an integer aspect ratio. Wavelength
ratios were $\Lambda_{o}/\Lambda_{z}= n/41$ with $n$ an integer.
Examples of final patterns of such simulations are shown in Fig.~\ref{fig:PWstab_inc} using the 
high order gradient-type inter-map coupling energy.
In all studied cases the final patterns are spatially regular.
The observed patterns are either fractured stripe patterns with two active modes 
Fig.~\ref{fig:PWstab_inc}\textbf{\sffamily{A,C,D,E}} 
or rPWC (two modes plus the corresponding opposite modes), see 
Fig.~\ref{fig:PWstab_inc}\textbf{\sffamily{B,F}}.
For a large ratio $k_{c,o}/k_{c,z}$ the patterns with two active modes are pinwheel 
free, see Fig.~\ref{fig:PWstab_inc}\textbf{\sffamily{A,C,D}}.
For smaller ratios $k_{c,o}/k_{c,z}$ the patterns are 
pinwheel rich, see Fig.~\ref{fig:PWstab_inc}\textbf{\sffamily{E}}.
The pinwheel density appears to exhibit a complex dependence on the wavelength ratio
since solutions switch back and forth between pinwheel-free fracture stripes and pinwheel-rich rPWCs
when the wavelength ratio is decreased from 1. Much larger domains than used in the current
simulations would be needed to simulate values intermediate to the wavelength ratios used here.
Our results, nevertheless, clearly establish that OD induced pinwheel stabilization
can occur also with detuned wavelengths.
They furthermore confirm that wavelength detuning does not by itself generates irregular stable
maps in the considered model.
\clearpage
\subsection{Higher feature numbers}\label{sec:ThreeMaps}
\begin{figure}[bt]
\begin{center}
\includegraphics[width=.75\linewidth]{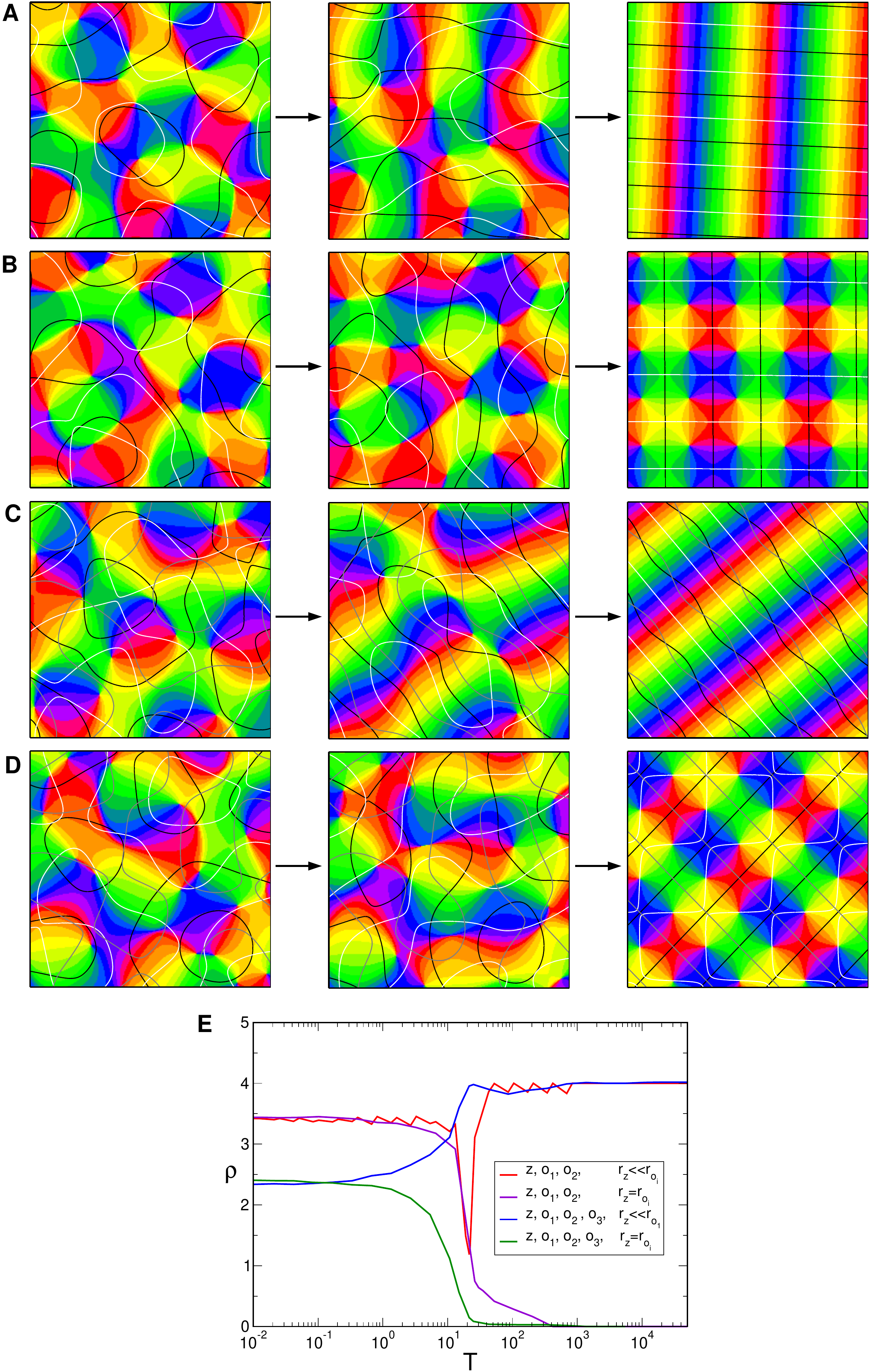} 
\end{center}
\caption{\textbf{Map interactions in higher feature dimensions.}
\textbf{\sffamily{A,B}} Map layout by interactions between three columnar systems 
($z(\mathbf{x}),o_1(\mathbf{x}),o_2(\mathbf{x})$). All maps are mutually coupled.
Superimposed on the OP map there are the borders of two real fields (black, white).
\textbf{\sffamily{A}} $r_z=r_{o_1}=r_{o_2}=0.1$
\textbf{\sffamily{B}} $r_z=0.01, r_{o_1}=r_{o_2}=0.1$.
\textbf{\sffamily{C,D}} Interactions with four columnar systems 
($z(\mathbf{x}),o_1(\mathbf{x}),o_2(\mathbf{x},t),o_3(\mathbf{x},t)$).
\textbf{\sffamily{C}} $r_z=r_{o_1}=r_{o_2}=r_{o_3}=0.1$.
\textbf{\sffamily{D}} $r_z=0.01, r_{o_1}=r_{o_2}=r_{o_3}=0.1$.
Superimposed on the OP map there are the borders the of three real fields (black, gray, white).
Left panel: initial conditions, middle panel: $T=10$, right panel: $T=T_f=5\cdot 10^4$.
\textbf{\sffamily{E}} Time evolution of the pinwheel density.
Parameters in all simulations: $\epsilon=2000, \gamma=0, \Gamma=22$, $128\times 128$ mesh.
}
\label{fig:ThreeFields}
\end{figure}

The inclusion of more feature dimensions into the dynamics can be performed in a 
similar fashion as in Eq.~(\ref{eq:DynamicsSHcoupled_ref}), Eq.~(\ref{eq:Energy_all}) as the 
geometric correlations between the different types of maps seem to be qualitatively 
similar \cite{Bonhoeffer,Das2,Grinwald_Dir,Sur5,Bosking3}.
To illustrate this we used the higher order gradient-type inter-map coupling with three 
and four maps which are mutually coupled. 
Whereas in the case of two maps the coupling energy is zero if the two stripe
solutions are perpendicular to each other the interactions between more maps can
potentially lead to a frustration as not all of the individual coupling energies can simultaneously vanish.
We used different, spatially irregular planform solutions as intitial conditions with initial pinwheel
densities $\rho \approx 3.4$ and $\rho \approx 2.4$.
Using the gradient coupling energy 
\begin{equation}
 U=U_1+U_2+U_3 =\epsilon_1\, |\nabla z \dotp \nabla o_1|^4+\epsilon_2\, |\nabla z \dotp \nabla o_2|^4
+\epsilon_3 \, |\nabla o_1 \dotp \nabla o_2|^4 \, ,
\end{equation}
no OD bias ($\gamma=0$), and equal coupling strengths $\epsilon_1=\epsilon_2=\epsilon_3=\epsilon$ 
we observed two types of stationary solutions, see Fig.~\ref{fig:ThreeFields}. 
In case where all bifurcation parameters were equal the OP map consisted of stripes.
Also the two real fields consisted of stripes, both perpendicular to the OP stripes i.e.
\begin{eqnarray}
 z(\mathbf{x})&=&Ae^{\imath \vec{k}_1 \cdot \vec{x}} \nonumber \\
o_1(\mathbf{x})&=&2B_1 \cos(\vec{k}_2 \cdot \vec{x}) \nonumber \\
o_2(\mathbf{x})&=&2B_2 \cos(\vec{k}_2 \cdot \vec{x}+\psi)\, , \quad \vec{k}_1 \cdot \vec{k_2}=0 \, .
\end{eqnarray}
The energy in this case is given by $U_1=U_2=0$, 
$U_3=\frac{B_1^4B_2^4\pi}{16}\left(18+16 \cos (2\psi)+\cos(4\psi)\right)$ 
which is minimal for $\psi=\pi/2$, i.e. the energy is minimized by shifting one 
real field by one quarter of the typical wavelength.
When the bifurcation parameter of the OP map was smaller than that of the 
two real fields we obtained PWC patterns, see Fig.~\ref{fig:ThreeFields}\textbf{\sffamily{B}}.
The pinwheels were arranged such that they are in the center of a square spanned
by the two orthogonal real fields and the resulting pinwheel density is $\rho=4$.
All intersection angles between iso-orientation lines and borders of the real fields
were perpendicular.
When extending the system by another real field we observed a similar behavior.
Figure~\ref{fig:ThreeFields}\textbf{\sffamily{C,D}} shows the stationary states of a complex field
coupled to three real fields.
In case of equal bifurcation parameters the stationary patterns were OP stripes,
perpendicular to stripe and meandering real patterns.
In case the bifurcation parameter of the OP map was smaller than the other bifurcation parameters
we again observed pinwheel crystallization. Note, that in this case all pinwheels were 
located at the border of one of the three real fields. 
To summarize,
pinwheel crystallization was only observed when the OP map is driven by the real field i.e.
when the OP amplitudes are small. 
In all observed cases the patterns were spatially perfectly periodic.
\clearpage
\begin{figure}[bt]
\begin{center}
\includegraphics[width=.9\linewidth]{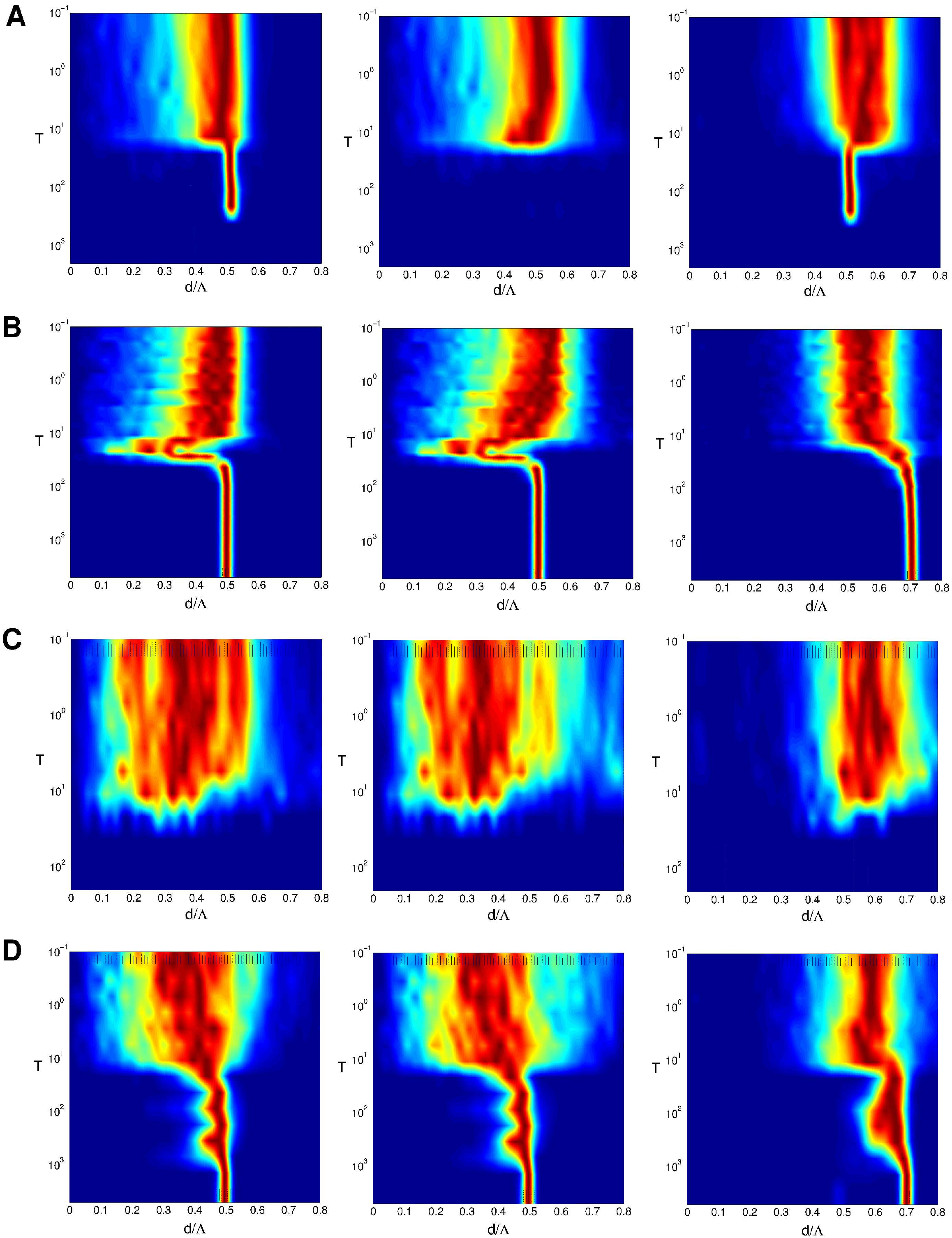} 
\end{center}
\caption{
\textbf{Distribution of nearest neighbor pinwheel distances in higher feature dimensions.}
Distance to the next pinwheel of arbitrary (Left), opposite (Middle), and
equal (Right) topological charge.
Parameters as in Fig.~\ref{fig:ThreeFields}.
\textbf{\sffamily{A,B}} Map layout by interactions between three columnar systems 
($z(\mathbf{x}),o_1(\mathbf{x}),o_2(\mathbf{x})$),
\textbf{\sffamily{A}} $r_z=r_{o_1}=r_{o_2}=0.1$,
\textbf{\sffamily{B}} $r_z=0.01, r_{o_1}=r_{o_2}=0.1$.
\textbf{\sffamily{C,D}} Interactions with four columnar systems 
($z(\mathbf{x}),o_1(\mathbf{x}),o_2(\mathbf{x},t),o_3(\mathbf{x},t)$),
\textbf{\sffamily{C}} $r_z=r_{o_1}=r_{o_2}=r_{o_3}=0.1$,
\textbf{\sffamily{D}} $r_z=0.01, r_{o_1}=r_{o_2}=r_{o_3}=0.1$.
}
\label{fig:ThreeFieldsDist}
\end{figure}
\clearpage
\section*{Discussion}
\textbf{Summary of results}\\
In this and the accompanying analytical study, we presented a dynamical systems approach 
to the coordinated optimization of 
maps in the visual cortex such as OP and OD maps. 
In part (I) we examined in particular such analytically obtained  
optima of various candidate energy functionals \cite{Reichl5}.
We calculated phase diagrams for different energy functionals showing that 
for strong inter-map coupling pinwheel crystals are optima of the system.
In the current study, we numerically analyzed the dynamics of one representative example
of these coordinated optimization
models. 
For this we focused on the high order gradient-type inter-map
coupling energy that can reproduce all qualitative relationships found experimentally between 
OP and OD maps, does
not suffer from potential OP map suppression, and has a relatively simple phase diagram near
the symmetry breaking threshold.
The main phenomenon characterizing the considered 
models, crystallization induced by coordinated optimization
and inter-map coupling was verified numerically.
This phenomenon 
was found to be robust to the influence of a weak backreaction of the OP map on the OD map, to detuning of 
the typical wavelengths, and was found to persist in models with higher feature space dimensionality.
We characterized the complex dynamics during crystallization and calculated various pinwheel
statistics.
The crystalline periodic layout of pinwheel-rich solutions persisted in all studied conditions.
Characterizing the behavior of transients, we found that spatially irregular transient states 
decayed relatively fast into locally ordered patterns during optimization.\\
When starting the optimization dynamics from small amplitude random initial conditions early 
OP patterns are essentially random orientation maps exhibiting a model insensitive, universal spatial 
organization throughout the initial emergence of orientation selectivity. 
As soon as orientation selectivity started to saturate the patterns typically reorganized 
towards one of a few crystalline spatial patterns. While pattern rearrangement can unfold 
over long timescales to achieve long-range spatial coherence, the key layout features of the 
crystalline target state were expressed after at most ten fold of the characteristic time of the 
initial growth of orientation selectivity. Similar behavior was also observed when starting 
near spatially irregular unstable fixed points of the orientation map dynamics. \\
\textbf{Dynamics during the optimization process}\\ 
Our study allows to examine the sequence of stages predicted for the maps under the assumption
of developmental optimization.
Our results consistently show that coordinated optimization models exhibit a complex
dynamics that persistently reorganizes maps over different timescales before
attractors or optimized ground states are reached.
Although our models exhibit various different attractor states none of the solutions 
converged directly to the final layout.
Rather, in all of them, a substantial rearrangement of maps after initial
symmetry breaking was found.
As can be predicted from symmetry \cite{Wolf3,Schnabel1}, 
at early stages of development maps must be spatially irregular if they develop
from weakly tuned random initial conditions. The average pinwheel density in these early maps
is bounded from below by the mathematical constant $\pi$ and the distributions of nearest 
pinwheel distances are continuous and broad.
In the models considered here,
these spatially irregular early states decay over a period of the 
order of the intrinsic timescale after saturation
of orientation selectivity.
This early phase of local crystallization rapidly leads to the occurrence of different spatial domains within 
the pattern, with a locally stereotypical periodic layout.
Even in the cases exhibiting the slowest decay of irregular patterns, this process
was complete after ten intrinsic timescales.
The slower dynamics that characterizes 
further development rearranges these domains to progressively align leading to a long-range ordered 
perfectly periodic crystalline array.
This reorganization of patterns lasts substantially longer than the intrinsic timescale.
The overall progression of states observed in our models has been found 
previously in numerous pattern forming
dynamics both highly abstract as well as in detailed ab initio simulations \cite{Cross,Manneville,Greenside}.
PWC solutions represent attractor states and
we found no other, spatially irregular, long-living states in the dynamics.\\
\textbf{Comparing the dynamics of coordinated optimization to stages of visual cortical development}\\
Does the observed rapid decay of irregular OP layouts into crystalline patterns speak 
against the biological plausibility of an optimization dynamics of the type considered here? 
Can one reasonably expect that a similar crystallization process could also unfold 
relatively rapidly during the development of the brain? Or is it more likely that what 
seems rapid in our numerical simulations would take very long in a biological 
network - potentially so long that the cortical circuitry has already lost its potential 
for plastic reorganization before substantial changes have occurred? 
To answer these questions it is important (1) to examine the fundamental 
timescales of the postnatal development of visual cortical circuits sub-serving 
orientation preference and ocular dominance, (2) to discuss how these timescales can be 
compared to the formal timescales that appear in dynamical models of map formation and 
reorganization and (3) to examine whether there is any biological evidence for a 
secondary rearrangement of orientation and ocular dominance maps subsequent to 
their initial establishment. \\
Because all reorganization processes predicted by our models occur after the 
saturation of single cell selectivities these questions are in fact essential to assess 
the explanatory power of the model. The predicted reorganization processes can only 
be reasonably compared to the course of biological development, if there is a period of 
ongoing cortical circuit plasticity that is substantially longer than the time needed to 
reach mature levels of single cell selectivity. The applicability of any dynamical model 
beyond the emergence of single cell selectivity would be fundamentally questioned if an 
arrested-development scenario would be adequate. Such an arrested-development scenario 
would imply that some random pattern emerging initially is effectively frozen by the 
termination of the period of developmental plasticity. 
In this case, viewing cortical development as an optimization process were to 
offer only very limited insight: Because the early emerging states are model 
insensitive \cite{Wolf3,Kadar,Wolf2,Schnabel1}
the same kind of frozen states would be predicted by interrupted optimization of very 
different energy functionals. 
This would have two distinct consequences: (1) biological observations on map 
structure would not contain much information to separate adequate from inadequate 
models of map formation and optimization and (2) none of these energy functionals 
could serve as a compact and specific description of the design principles ruling cortical 
functional architecture. \\
Let us first discuss the duration of the process of establishing mature “saturated” levels 
of stimulus selectivity from weakly selective initial conditions from a theoretical perspective. 
All models for the development of visual cortical functional selectivity from an 
unselective or weakly selective initial condition are known to exhibit a distinct intrinsic 
time scale for the emergence of stimulus selectivity in individual 
neurons \cite{Malsburg2,Ritter,Swindale2,Miller3,Miller,Swindale6}.
In the abstract order parameter models examined here this time scale is set by the inverse of the 
maximum eigenvalue $r_z$ . It is important to note that this time scale represents an 
effective parameter describing a collective circuit property and that this parameter is not 
rigidly related to any particular cellular or synaptic timescale such as e.g. the characteristic 
time required for the expression of LTP, spine growth, homeostatic plasticity or other functional or 
morphological synaptic changes. Theoretical studies of microscopic models in which the 
effective time scale was explicitly calculated have established that this time scale 
depends (1) on the ensemble of activity patterns driving development and also (2) on characteristics 
of the local cortical circuits \cite{Scherf,Miller3,Miller6}. 
For instance in a representative, analytically solvable microscopic model for the 
emergence of ocular dominance patterns the maximum eigenvalue is given by 
$$\lambda_{max}=-\frac{\sigma^2}{\tau_{syn}}+\frac{1}{\tau_{syn}}\int d^2x \left( C_L(x)+C_R(x)-2C_{LR}(x) \right)$$
where $\tau_{syn}$ is the characteristic time scale for synaptic changes, 
$\sigma$ measures the spatial extend of co-activated neuron groups in the model cortex and 
$C_L$, $C_R$ and $C_{LR}$ are auto- and cross-correlation functions of the activity patterns in the 
left and right eye layers of the LGN \cite{Scherf}. 
While the effective time scale for the emergence and saturation of selectivity is expected not 
to be faster than the fundamental processes of synaptic change, they can certainly be substantially slower. \\
The best way to obtain an estimate for the intrinsic time scale of map dynamics in the developing 
visual cortex is thus to consider the empirically observed duration of the respective processes 
under normal conditions. For orientation selectivity in the primary visual cortex this information 
has been obtained conclusively for cat and ferret. In both species orientation selectivity is established 
starting from an initial condition in which cells are only weakly orientation biased and it takes 
a period between a few days and at most one week to reach mature levels
of single cell orientation selectivity 
\cite{Albus,Fregnac,White,White3,Chapman,Crair}. 
It is noteworthy that similar time scales are also sufficient for substantial morphological changes 
of thalamo-cortical axonal structure \cite{Antonini}. 
Even in anesthetized animals, visual cortical preferences for stimulus orientation or 
direction, or preference for one eye all can undergo substantial activity induced changes 
within a few hours 
\cite{Fregnac2,Fregnac3,Fregnac4,Huebener,Fitzpatrick}.
The characteristic time scale for the emergence and maturation of single cell stimulus selectivities 
can thus be constrained with good confidence to be of the order of at least a few hours 
and at most one week. \\
The second time scale that needs to be estimated empirically is the duration of the period over 
which juvenile levels of developmental plasticity in visual cortical circuits are maintained. 
This period can be narrowed down by considering the period of susceptibility for 
modification of cortical neuronal preferences by changes of visual experience. 
For this, it is to be kept in mind that statements on the duration of such periods of 
susceptibility invariably depend on a particular type of modification of visual 
experience and specific circuit involved 
\cite{Fregnac,Movshon,White3}.
The temporal extend of a period of plasticity has been addressed most comprehensively 
for the modification of ocular dominance by monocular deprivation in the cat visual cortex. 
Ocular dominance plasticity is mediated by thalamo-cortical projections. 
Thalamo-recipient neurons in the cat visual cortex are also the first neurons in the thalamo-cortical 
visual pathway which exhibit orientation selectivity and thalamo-cortical connections are 
thought to define the preferred orientation of cortical columns in the cat (see e.g. \cite{Miller5}). 
It is thus plausible to assume that ocular dominance plasticity is probing the plasticity of 
circuitry underlying both orientation preference and eye dominance. 
It is conceivable in principle that the features of thalamo-cortical circuitry that define the 
preferred orientation - although mediated by the same synapses - are much less susceptible to 
modification than those defining eye dominance. 
The emergence of distinct receptive field features appear to be differentially sensitivity to the 
ablation of particular synaptic molecular components, indicating that the emergence of 
orientation selectivity requires a temporally more precise synaptic transmission than 
ocular dominance segregation and plasticity \cite{Fagiolini}. 
An explicit mechanism, however, that could terminate plasticity of orientation preference 
while preserving it for ocular dominance, has neither been identified experimentally nor theoretically. 
On the contrary, Fregnac and coworkers in classical microstimulation studies found that the 
very same stimulation strategy that was effective in changing the ocular dominance of visual 
cortical neurons during the ocular dominance critical period was also effective in inducing changes 
of their preferred orientations \cite{Fregnac2,Fregnac3,Fregnac4}.
In addition, reversal of
monocular deprivation in area 17 (but not area 18) can induce the formation of different 
preferred orientations in the left and right eye receptive fields of a neuron 
\cite{Blakemore,Movshon2,Goedecke}.
Overall the available evidence in the cat thus supports the view that ocular dominance 
plasticity probes the plasticity of neuronal circuitry critical to both ocular dominance 
and orientation preference. 
We will summarize the current evidence that the orientation preference of cortical neurons 
in fact changes during the critical period for monocular deprivation in the 
section 'Juvenile plasticity can support an ongoing reorganization of orientation preference 
and ocular dominance' below.\\
How long does the period of juvenile ocular dominance plasticity last? 
In the field of visual cortical development, there is a broad and well established 
consensus that the period of susceptibility to monocular deprivation in the cat lasts 
for several months of postnatal live \cite{HubelWiesel5,Blakemore,Olson,Fregnac,Jones}. 
The primary visual cortex of the cat is maximally susceptible to monocular 
deprivation in kitten of four weeks of age \cite{HubelWiesel5,Blakemore,Olson}.
The initial establishment of orientation preference and ocular dominance maps in 
kittens occurs around the time of eye opening in the second postnatal week \cite{Crair,Albus}.
Maximal sensitivity is thus reached two weeks after the emergence of maps and the 
onset of normal vision.  After the first postnatal month susceptibility to monocular 
deprivation was found to gradually decline back to levels comparable to those present 
at the onset of vision and initial map emergence. The closure of the period of 
developmental plasticity was estimated by several studies to occur between the 
14th and 18th postnatal week \cite{HubelWiesel5,Blakemore,Olson}.
This is 12 to 16 weeks after single neurons first exhibit adult like levels 
of orientation selectivity and eye dominance. 
Later studies, found that substantial modifications of responses could be induced 
even up to the 35rd postnatal week \cite{Jones}.\\ 
The available data thus indicate a clear separation of timescales between the duration of the 
period of developmental plasticity and the characteristic time scale for the emergence of 
stimulus selectivity and maps. The duration of the relevant period of developmental plasticity 
is at least 10 times longer (e.g. 12 weeks vs. 1 week) and potentially more than three 
orders of magnitude longer (32 weeks vs. 6 hours) than the characteristic time scale 
for the emergence of stimulus selectivity and maps. This estimate rests on the idealizing 
assumption that the characteristic time scale of map dynamics remains essentially constant 
over the duration of the critical period. An assumption that is not easy to check and not 
likely to be exactly fulfilled. If the time constant was to rapidly increase after the initial 
generation of the maps, the separation of time scales would be less pronounced. 
However, the susceptibility to ocular dominance plasticity, in fact, increases after 
eye opening and initial map formation and returns to levels as low as during initial map
establishment only at the end of the critical period \cite{Blakemore,Olson}.
A higher degree of plasticity during the period of juvenile plasticity would be rather 
consistent with a more speedy process i.e. a reduced time constant than with an 
increased, i.e. slowed, down time constant. This observation thus suggests that our 
estimate rather under estimates than over estimates the separation of the two time scales. \\
We emphasize that we are certainly not the first to point out this substantial 
separation of time scales. 
To our best knowledge, Jack Pettigrew was the first to highlight and 
discuss the issues and question raised by this fact of cortical circuit 
development in his essay entitled \textit{The paradox of the critical 
period for striate cortex} (\cite{Pettigrew}, see discussion in \cite{Kaschube5}). 
Specifically with respect to modeling 
the dynamics of functional cortical architecture during early development this aspect has 
been considered previously particularly in \cite{Wolf3,Wolf1,Kaschube6,Kadar}.\\ 
\textbf{When does a dynamical model successfully explain adult functional cortical architecture? }\\
In light of the above considerations it appears of limited value to compare maps 
from a simulation obtained when selectivity first reaches mature levels to biological patterns 
present in the adult cortex. Maps in the adult visual cortex of the cat have been subject to 
more than ten weeks of ongoing plasticity. They are therefore better viewed as dynamic 
equilibria that are largely maintained under a continuous process of ongoing activity-driven 
synaptic turnover. Current experimental evidence, nevertheless, indicate that the maps 
emerging initially over the first days of normal vision exhibit many layout properties that 
are preserved throughout the juvenile period of plasticity and into adulthood 
\cite{Kaschube6,Chapman}.
Taking the long duration of the period of juvenile plasticity into account, this is 
likely to mean that these properties have been actively maintained by an ongoing dynamics. \\
Our current results as well as many prior theoretical studies (reviewed in \cite{Kaschube6}, supplement) 
clearly demonstrate that the requirement to generate and maintain a realistic column layout is a much 
more selective criterion for the identification of appropriate candidate models than 
the mere ability to initially generate “good looking maps”. 
It is thus a more stringent test of a dynamical model’s explanatory power to compare 
the maps obtained at later stages to the biologically observed functional organization 
of the visual cortex. Our estimates suggest that it would be reasonable to require of a 
biologically plausible model that states which resemble the adult functional 
architecture occur in time between one and three orders 
of magnitude later than the maturation of average selectivity. Using this criterion, the states 
observed in our simulations in fact suggest 
that the considered models are not capable of explaining the biological organization in a 
satisfying fashion: Patterns observed between 10 and 1000 intrinsic timescales are 
dominated by crystalline local arrangements that are distinctly different from the spatially irregular
layout of orientation maps observed in both juvenile and adult visual cortex. \\
\textbf{Juvenile plasticity can support an ongoing reorganization of orientation preference and ocular dominance }\\
The comparison of patterns at between 10 and 1000 intrinsic timescales suggests that the 
maintained plasticity that is ongoing at least until the end of the critical period has features 
not present in the models examined here. Biologically, this conclusion would be completely
compulsory, if there was direct evidence that the period of juvenile plasticity in fact included a 
substantial reorganization of visual cortical functional organization after the initial emergence 
of single cell selectivity. This would constitute direct evidence that the initially emerging set 
of stimulus preference is in fact not frozen and that the duration of the period of juvenile 
plasticity is of sufficient duration to allow a secondary reorganization of the maps. \\
Recently, evidence that ongoing plasticity enables ongoing pattern reorganization 
has started to accumulate both for species exhibiting a columnar functional architecture 
as well as for the rodent typical non-columnar salt-and-pepper organization of stimulus preferences 
\cite{Keil,Rochefort,Kaschube5}.
For cat visual cortex, Kaschube and coworkers have demonstrated that the spatial 
organization of orientation columns in striate cortex is progressively reorganized between 
the sixth and the 14th postnatal week such that the organization of orientation columns 
that are reciprocally connected to extra-striate visual cortex and contra-lateral hemisphere 
striate cortex are better matched \cite{Kaschube5}. A second line of evidence is related to 
the fact that the surface area of cat striate cortex substantially increases postnatally
\cite{Duffy,Keil,Loewel11,Villablanca}. 
The spatial periodicity of both orientation as well as ocular dominance columns, however, remains 
basically unaffected during this period 
\cite{Keil,Loewel12,Kaschube5}.
Keil and coworkers have reported evidence that this areal growth 
in the presence of maintained mean column spacing induces a characteristic 
spatial reorganization of the layout of ocular dominance columns within cat 
striate cortex \cite{Keil}. 
Growth related rearrangement of orientation columns had been suggested 
previously from observations on a smaller data set from juvenile macaques \cite{Blasdel6}.\\
Perhaps the most striking demonstration that the functional preferences of visual 
cortical neurons can reorganize over long time scales during the period of juvenile plasticity 
has emerged from studies of the mouse visual cortex. In the mouse, as in cat, visual 
cortical neurons first develop orientation selectivity around the time of first eye 
opening in the second postnatal week \cite{Wang,Rochefort}.
Mimicking the developmental time course in the cat, also the duration of the 
period of juvenile plasticity in the mouse is quite long and extends beyond the 
third postnatal month \cite{Stryker4,Loewel10}.
At a duration of more than 10 weeks, it is thus substantially longer than required for the expression of
adult-like single neuron selectivities.  
Wang and coworkers recently demonstrated that neurons in the binocular 
segment of mouse visual cortex in fact change their preferred orientations 
during this period \cite{Wang}. 
Neurons in the binocular segment of mouse striate cortex were found to 
first exhibit widely different preferred orientations in the left and right eye. 
The two different preferred orientations then underwent secondary reorganization 
and became matching at an age of 5 weeks postnatally \cite{Wang}. 
Such late changes might not be restricted to the binocular segment of the mouse visual cortex. 
In the monocular segment, Rochford and coworkers documented substantial changes 
in the complement of preferred orientations and preferred directions represented 
between eye opening and the first postnatal month \cite{Rochefort}. 
An adult-like composition of orientation selectivities was not found before 
the end of the second postnatal month \cite{Rochefort}. It might be interesting 
to note that a substantial long-term reorganization of cortical preferences has 
also been recently demonstrated in the rat barrel cortex \cite{Kremer}. 
Kremer and coworkers found that preferences for the direction of whisker 
deflection reorganize over the course of the first three postnatal months. 
Long-term reorganization might thus potentially constitute a general 
feature of sensory cortical representations.\\
Experimental evidence thus clearly supports that the circuits underlying ocular 
dominance and orientation selectivity remain in a state of flux for weeks and 
months after the initial emergence of visual responsiveness and stimulus selectivity.  
We conclude that the model considered in the present study is - in its present form 
incapable of reproducing the maintenance of visual cortical architecture of 
primates and carnivores in a convincing manner. 
This conclusion presumably also applies to all models examined in the 
accompanying analytical study. In none of them did we find a stable state 
or minimum energy pattern that would reproduce the layout of orientation 
columns in the visual cortex. It is thus predicted for all of them that spatially 
irregular states, if formed transiently, will decay into biologically unrealistic 
crystalline patterns. How such models could be modified to better reproduce the 
observed spatially irregular layout of cortical columns will be discussed subsequently in
the section 'Generality of pinwheel crystallization'. \\
\textbf{Conditions for pinwheel stabilization}\\
The analytical results presented in part (I) showed that in several models OD stripes are not 
able to stabilize pinwheels near symmetry breaking threshold and for only one real-valued scalar field. 
This result appeared to be insensitive of the specific type of inter-map interaction \cite{Reichl5}.
Our numerical results show that this result is also insensitive to 
a detuning of typical wavelengths.
For different ratios of the typical wavelengths of OP and OD pinwheel-rich patterns either decay into 
pinwheel-free OP stripes or patterns with OP fracture lines when interacting with OD stripes.
These findings support the conclusion that in models for the joint optimization of OP and OD maps
a patchy OD layout is important for pinwheel stabilization
by crystallization. 
One should note, however, that sufficiently far above threshold dislocations 
in the OD map can become frozen \cite{Boyer}.
Via inter-map coupling these dislocations could prevent pinwheels to completely annihilate
and thus lead to local pinwheel stabilization near OD dislocations.\\
In the current study we also generalized  
our dynamical systems approach to to include any additional number of columnar systems. 
One reason to consider additional visual cortical maps originates from the
finding that the removal of the OD map in experiments does not completely destabilize pinwheels \cite{Sur2}.
Moreover, in tree shrews, animals which completely lack OD columns, OP maps contain 
pinwheels and exhibit a pinwheel arrangement essentially indistinguishable from species
with columnar OD segregation \cite{Bosking}.
This might reflect the influence of additional columnar systems like spatial frequency columns 
that can be expected to interact with the OP map in a similar fashion as OD columns \cite{Bonhoeffer}.
From a theoretical perspective, one might suspect that couplings between more than two
systems that promote a mutually orthogonal arrangement are harder to satisfy the more maps
are considered. In principle this could lead to the emergence of irregular patterns by frustration. 
In numerical simulations we examined coordinated optimization with three and four columnar systems.
In these cases pinwheel stabilization is possible even without an OD bias. 
The resulting stationary OP patterns are, however, still either stripes or PWC solutions.
For more than two feature maps, asymmetry of one feature dimension is thus not 
a necessary condition for pinwheel stabilization
by coordinated optimization.\\ 
We also characterized the dynamics of pinwheel crystallization from pinwheel-free initial conditions.
With the analytical approach presented in part (I) we were able to show that pinwheel-rich solutions 
correspond to the energetic ground state of our models for large inter-map coupling \cite{Reichl5}.
This can be reflected by simulations in which pinwheels 
are created even when starting from an initial OP stripe pattern. 
Assessing pinwheel creation from pinwheel-free initial conditions could more generally
serve as a simple test for the existence of a pinwheel-rich attractor state 
in models of OP development that can
be applied to models of arbitrary complexity.
One should note, however, that the production of pinwheels from a pinwheel-free initial condition
provides only a sufficient but not necessary criterion to verify the existence
of a pinwheel-rich attractor state. 
This criterion may be violated if pinwheel-free and pinwheel-rich
attractor states coexist.
Nevertheless, the pinwheel production criterion can be used to demonstrate 
that pinwheels are not just a remnant of random initial conditions.\\
\textbf{Generality of pinwheel crystallization}\\ 
Pinwheel crystals have been previously found in several abstract \cite{Koulakov,Kadar} as well as
in detailed synaptic plasticity based models \cite{Malsburg2,Malsburg,Min}.
Remarkably, in a model of receptive field development based on a detailed dynamics of synaptic connections
the resulting OP map showed a striking similarity to the 
hPWC presented above, compare Fig.~(\ref{fig:PWdistance_last}) and \cite{Reichl5,Malsburg}.
These observations indicate that pinwheel crystallization is not an artefact of the highly idealized
mathematical approach used here.
In fact, the first OP map predicted ever by a synaptically based self-organization model presented by von
der Malsburg in 1973 exhibited 
a clearly hexagonal column arrangement \cite{Malsburg2}.
Von der Malsburgs calculations as well as those presented in \cite{Malsburg} utilized a
hexagonal grid of cells that may specifically support the formation of hexagonal patterns.
Our numerical and analytical results clearly demonstrate that patterns of hexagonal symmetry
do not critically depend on the use of a hexagonal grid of cells.
As our simulations can generate hexagonal pattern also for square lattices of cells
it would be instructive to revisit von der Malsburgs model implemented for other
grids both of square symmetry as well as for irregular positions of cells.
In our study, we examined whether the non-crystalline layout of visual cortical maps
could result from a detuning of OP and OD wavelengths. 
However, while destabilizing 
hPWC solutions, wavelength detuning leads to spatially perfectly regular solutions in all studied cases.
This suggests that a spatially regular layout is  not an exceptional behavior in models of the coordinated
optimization of visual cortical maps that would require fine tuning of parameters.\\
The reason for the strong differences to experimentally observed maps might thus be the presence 
of biological factors 
neglected in the models examined here. 
Such factors might be a greater distance from the pattern formation threshold, 
different kinds of biological noise, or the presence of long-range neuronal interactions. 
Vinals and coworkers demonstrated for the case of stripe patterns that a Swift-Hohenberg 
model far from the bifurcation point can exhibit stable disclination defects \cite{Boyer}. 
Although the results indicate only a spatially sparse set of stabilized defects it will be 
interesting to examine whether this also applies to the case of multidimensional coupled 
Swift-Hohenherg models and to establish which properties model solutions develop very 
far from the bifurcation point. 
Theoretically, it is well understood that in principle so called 'nonadiabatic effects' can 
induce the pinning of grain boundaries in 
pattern forming media \cite{Pomeau3,Malomed,Bensimon,Vinals4}.
In one spatial dimension and for models with several interacting order parameter 
fields similar mechanisms may even lead to the emergence of spatially chaotic solutions 
\cite{Jacobs,Coullet,Coullet2}. These studies suggest to examine
whether spatial incommensurability far from threshold can induce spatially chaotic
patterns in one and two dimensional coordinated optimization models. Such studies may uncover
a completely novel scenario for explaining the emergence of spatially irregular states in models 
of cortical map optimization.\\
A second interesting direction will be the inclusion of frozen spatial disorder 
in models for the self-organization of multiple cortical maps. 
Such disorder could represent a temporally fixed selectivity bias that favors 
particular feature combinations at different position in the cortical sheet. 
For orientation preference, the proposals of Waessle and Soodak recently 
revisited by Ringach and coworkers that retinal ganglion mosaics might constrain 
and seed orientation column patterns would represent a specific 
mechanisms for such a fixed local bias \cite{Ringach2011,Soodak,Wassle,Ringach2,Ringach3}. 
Experimental evidence that retinal organization can impose local biases 
was revealed by Adams and Horton’s finding that the pattern of retinal blood 
vessels can specifically determine the layout of ocular dominance columns in 
squirrel monkey visual cortex (\cite{Horton4,Horton6}, for a modelling study see also \cite{Goodhill10}). 
Spatial disorder terms in an equation can also be designed to model randomness 
in the interactions between neurons at different positions. 
This might result from heterogeneities in lateral interactions within the cortical sheet. 
In particular this later type of modification has been examined in simple examples of 
order parameter equations 
and was found to qualitatively change the type of the bifurcation and the nature of the 
unstable modes \cite{Zimmermann,Zimmermann2,Zimmermann3,Pomeau2}. It 
will be important to investigate how different types of spatial disorder 
modify the behavior of models derived from biologically meaningful energy functionals. 
We hope that for such studies of the influence of 'biological noise' a thorough 
understanding of the properties of perfectly homogeneous and isotropic systems as achieved 
here will provide a solid basis for disentangling the specific contributions of randomness 
and self-organization. \\
Finally, a third promising direction for modifying the type of models considered here 
is the inclusion of long-ranging intra-cortical interaction in the equations for the individual 
order parameter fields. The impact of long-ranging intra-cortical interactions has been 
studied previously both with respect to the properties of patterns emerging during the 
phase of initial symmetry breaking \cite{Shouval,Bartsch,Bednar,Bednar3}
as well as 
for its influence on long-term pinwheel stabilization and pattern selection \cite{Wolf1,Wolf2,Kaschube3}.
Models for orientation maps that include orientation-selective long-ranging interactions 
exhibit a good quantitative agreement of both attractors and transient states to the 
biological organization of orientation preference maps in the visual 
cortex \cite{Kaschube6}. Including orientation-selective long-ranging 
interactions in models for the coordinated optimization of multiple cortical maps 
could provide a transparent route towards constructing improved models for the 
coordinated optimization of column layouts matching the spatial structure of orientation maps. \\
\textbf{Experimental evidence for hexagonal orientation column patterns}\\
Recently, Paik and Ringach have argued that a roughly hexagonal arrangement of iso-orientation
domains would provide evidence for a defining role of retinal ganglion 
cell mosaics for the spatial arrangement of orientation columns \cite{Ringach2011,Ringach2,Ringach3}.
It is interesting to consider this claim in view of our results as well as in view of the wealth
of activity-dependent models that predict hexagonal arrangements 
irrespective of the arrangement of retinal ganglion cells \cite{Reichl4,Wolf1,Wolf2,Malsburg,Cowan2,Ernst}.
Since all of these distinct models are known to generate hexagonal arrays of 
orientation columns it seems questionable to view evidence for a hexagonal arrangement as evidence for a
particular activity-independent mechanism.
Our characterization of the dynamics of crystallization, however, enables to identify
more selective predictions of a retinal ganglion cell mosaic based formation
of hexagonal iso-orientation domains.
If the pattern of OP columns is seeded by retinal ganglion cell mosaics, as 
initially proposed by Soodak \cite{Soodak} and recently re-articulated 
by Paik and Ringach \cite{Ringach2011}, hexagonal structures should be detectable from the very
beginning of development, i.e. already at stages when orientation selectivity is still increasing.
Hexagonal PWCs in self-organizing models, in contrast, form from an initially irregular 
and isotropic state. Thus the time dependence of hexagonal-like column arrangements
can distinguish in principle between self-organized as opposed to retinal ganglion cell mosaic imprinted
hexagonal arrangements.
As we found in all models examined hexagonal arrangements are frequently of rather high
pinwheel density of about 5 pinwheels per hypercolumn.
Also the hexagonal pattern constructed by Paik and Ringach appear to exhibit relatively high pinwheel
density of $\rho \approx 3.5$. Thus both theories appear inconsistent with observed pinwheel densities.
A mixed scenario in which retinal ganglion cell mosaics seed the initial
pattern of iso-orientation domains and later activity-dependent refinement drives a 
rearrangement of OP maps towards the experimentally observed design therefore 
predicts a substantial degree of net pinwheel annihilation.
Kaschube et al. presented evidence for essentially age independent pinwheel densities
in ferret visual cortex starting between 5 to 20 postnatal weeks. 
No indication of substantial pinwheel annihilation is visible in this data \cite{Kaschube6}.
One should note that in ferret visual cortex orientation columns first arise
in the fifth postnatal week \cite{Chapman}.
The relation of the analysis by Paik and Ringach and the statistical laws described by Kaschube
and coworkers ask for further analysis and comparison.\\
\textbf{A hierarchy of visual cortical maps}\\
In the current studies we focused on a particular hierarchy of visual cortical maps. 
In the analytical calculations and most simulations the
OD map was assumed to be dominant which corresponds to a choice of 
control parameters that satisfy $r_o \gg r_z$.
That maps form a hierarchy under such conditions can be seen from the limiting case
in which inter-map interactions become effectively unidirectional.
In this case the dynamics of the OP map is influenced by OD segregation while the OD dynamics
is effectively autonomous. 
This limit substantially simplifies the analysis of map-interactions and the 
identification of ground states.
The effect of a backreaction on the OD map can be studied within the presented approach
either by solving amplitude equations numerically or by solving the full field dynamics.
We observed that, although the presented optima persist, with increasing backreaction
on the OD map the minimum inter-map coupling strength necessary for the stability
of hexagonal pinwheel crystals increases.
By solving the full field dynamics numerically we confirmed this conclusion.
In the presented numerical simulations the backreaction, however, was relatively small.
The simulations nevertheless establish that our results are not restricted to the limit $r_z/r_o \rightarrow 0$
i.e. that this regime does not represent a singular limit.
A comprehensive analysis of the effect of strong backreaction is  
beyond the scope of the current study.
One should note that the decoupling limit $r_z/r_o\rightarrow 0$ does not lead to completely 
unrealistic OD patterns. 
In particular, compared to the architecture of macaque visual cortex the uncoupled OD 
dynamics has stationary patterns which
qualitatively well resemble the layout of observed OD maps. 
Macaque primary visual cortex appears to exhibit essentially three different kinds of OD patterns:
Fairly regular arrays of OD stripes in most of the binocular part of the visual field representation,
a pattern of ipsilateral eye patches in a contralateral background near the transition zone to the monocular
segment and of course a monocular representation in the far periphery.
These might correspond to the three fundamental solutions of the OD equation: 
stripes, hexagons, and a constant solution, which are
stable depending on the OD bias (see Figure (16) of \cite{Reichl5}).
In cat visual cortex the observed OD layout is patchy \cite{LeVay2,Shatz,Shatz2,Loewel14,Kaschube2}. 
\\
\textbf{Modeling areal borders and experimentally induced heterogeneities}\\
Viewed from a formal perspective, 
the presented theoretical approach offers also convenient ways to model the impact of 
spatial inhomogeneities in the
visual cortex on OP map structure.
For this purpose, the co-evolving field does not represent a feature map but would be
designed to describe a real or artificial areal border or a 
disruption of local circuitry.
To this end, the OP map would be coupled, using low order 
coupling energies, to a fixed field describing the areal border such that its values are for instance
one inside and minus one outside of the area with a steep gradient interpolating between the two. 
Outside the areal border
a strong coupling to such a field can lead to complete suppression of orientation selectivity.
Using a gradient-type inter-map coupling energy inter-map coupling can also be used to favor a 
perpendicular intersection of iso-orientation lines with the areal borders as observed 
in some experiments \cite{Bosking,Loewel9}.
Artificial heterogeneities and areal borders have been induced by local ablation
or other local surgical interventions \cite{Loewel9}. 
Viral approaches such as the silencing of cortical regions by transfection with hyperpolarizing
ion-channels now make it possible to impose them
with minimal intervention and potentially in a reversible fashion \cite{Johns,Slimko}.\\ 
\textbf{Conclusions}\\
The presented models for the coordinated optimization of maps in the visual cortex,
that were studied analytically in part (I) and numerically in part (II),
lead in all studied conditions to spatially perfectly regular energy minima.
In local regions on the order of a few hypercolumn areas, column layout rapidly
converges to one of a few types of regular repetitive layouts.
Because of this behavior the considered models cannot robustly explain 
the experimentally observed spatially 
irregular common design of OP maps in the visual cortex.
As expected from these qualitative differences all pinwheel statistics
considered, exhibit strong quantitative deviations from the experimentally observed values.
These findings appear robust with respect to finite backreaction, detuning of characteristic 
wavelengths, and the addition of further feature space dimensions.
Recent work demonstrated that the spatially irregular pinwheel-rich layout of pinwheels and orientation columns
in the visual cortex can be reproduced quantitatively by models that represent only the orientation
map but include long-range interactions \cite{Kaschube6,Wolf2,Wolf1}.
In the visual cortex of species widely separated in mammalian evolution we
previously found virtually indistinguishable layout rules of orientation columns that
are quantitatively fulfilled with a precision of a few percent \cite{Kaschube6}.
In view of these findings the current results suggest that models in which pinwheel stabilization
is achieved solely by coordinated optimization 
and strong inter-map coupling are not promising candidates for explaining visual cortical architecture.
In order to achieve a quantitatively more viable coordinated optimization theory one 
might consider taking additional 'random' factors into account.
One should however not conclude that coordinated optimization does not shape visual cortical
architecture.
Using the general approach developed here
it is possible to construct a complementary type of models in which the complex OP map is dominant.
Such models, using non-local terms in the energy functional of the OP map, can be
constructed to reduce to the 
model in \cite{Kaschube6,Wolf2,Wolf1} in the weak coupling limit.
Because long-range interaction dominated models can reproduce the spatially irregular layout of OP
maps, one expects from such models a better reproduction of the observed architecture for weak coupling.
Alternative scenarios might emerge from the inclusion of quenched disorder or very far from the 
pattern formation threshold.
Because of its mathematical transparency and tractability the approach developed in the present 
studies will provide powerful tools for examining to which
extend such models are robust against coupling to other cortical maps
and to disentangle the specific contribution of coordinated optimization
to visual cortical architecture.
\section*{Methods}\label{sec:MM}
\subsection*{Tracking and counting pinwheels}\label{sec:PWkinematics_part3a}
During the evolution of OD and OP maps we monitored the states from the initial
time $T=0$ to the final time $T=T_f$ using about 150 time frames.
To account for the various temporal scales the dynamics encounters the time frames
were separated by exponentially increasing time intervals.
Pinwheel centers were identified as the crossing of the zero contour lines of the
real and imaginary parts of $z(\mathbf{x})$. During time evolution we tracked
all the pinwheel positions and, as the pinwheels carry a topological charge, we divided the
pinwheels according to their charge.
The distribution of pinwheel distances indicates 
the regularity and 
periodicity of the maps. Therefore we calculated the minimal distance between pinwheels, 
measured in units of the column spacing $\Lambda$ during time evolution.
In simulations we used periodic boundary conditions.
In order to correctly treat pinwheels close to map borders we periodically 
continued the maps.
Nearest neighbors of pinwheels are thus searched also in the corresponding periodically continued maps.\\
The rearrangement of OP maps leads to annihilation and creation of pinwheels in pairs.
Between two time frames at $T_i$ and $T_{i+1}$ we identified corresponding pinwheels if their
positions differed by less than $\Delta x=0.2\Lambda$ and carry
the same topological charge. If no corresponding pinwheel was found within $\Delta x$
it was considered as annihilated. If a pinwheel at $T_{i+1}$ could not be assigned to one 
at $T_i$ it was considered as created.
We define the pinwheel creation $c(t)$ and annihilation $a(t)$ rates per hypercolumn as
\begin{equation}\label{eq:PWcreaPWannihi_part3a}
c(t)=\frac{d N_c}{\Lambda^2 dt},\quad  a(t)=\frac{d N_a}{\Lambda^2 dt}\, ,
\end{equation}
where $N_c$ and $N_a$ are the numbers of created and annihilated pinwheels.
Creation and annihilation rates were confirmed by doubling the number of time frames.\\
To what extend are the pinwheels of the final pattern just rearrangements of pinwheels at
some given time $T$?
To answer this question for a given set of pinwheels at an initial time $T=T^*$ we further 
calculated the fraction $s(t)$ of those pinwheels surviving until time $T$. 
Finally, the fraction of pinwheels present at time $T^*$ that survive up to the final time $T=T_f$ 
is given by $p(t)$.

\subsection*{Numerical integration scheme}\label{sec:App_IntegrationScheme}
As the Swift-Hohenberg equation is a stiff partial differential equation we used a fully 
implicit integrator \cite{Brown}. Such an integration scheme avoids numerical instabilities
and enables the use of increasing step sizes when approaching an attractor state.
The equation 
\begin{equation}
\partial_t \,z(\mathbf{x},t)=\hat{L} z(\mathbf{x},t)
-N[z(\mathbf{x},t)] \, , \quad \hat{L}=r-\left(k_c^2+\Delta\right)^2\, ,
\end{equation}
is discretized in time. Using a Crank-Nicolson scheme this 
differential equation is approximated by the nonlinear difference equation
\begin{equation}
\frac{z_{t+1}-z_t}{\Delta t} = \frac{ \left(
\hat{L} z_{t+1}+N[z_{t+1}] \right)+\left(\hat{L} z_t
+N[z_t] \right)} {2}.
\end{equation}
This equation is solved iteratively for $z_{t+1}$ with the help of the Newton method
which finds the root of the function
\begin{equation}
G[z_{t+1}]=\left(-\hat{L}+\frac{2}{\Delta t} \right) z_{t+1}
-N[z_{t+1}]-\left(\left(\hat{L}+\frac{2}{\Delta t} \right)z_t
+N[z_t] \right)\, .
\end{equation}
The field $z(\mathbf{x})$ is discretized. For a grid with $N$ meshpoints 
in the $x$-direction and $M$ meshpoints in the $y$-direction this leads to an $M\times N$
dimensional state vector $\mathbf{u}$. Discretization
is performed in Fourier space.
The Newton iteration at step $k$ is then given by
\begin{equation}\label{eq:App_NewtonMethod}
DG(u^k) \Delta \mathbf{u}^k=-G(\mathbf{u}^k), \quad \mathbf{u}^{k+1}=\mathbf{u}^k+\Delta \mathbf{u}^k \, ,
\end{equation}
with $DG$ the Jacobian of $G$.
Instead of calculating the matrix $DG$ explicitly a matrix free method
is used, where the action of the matrix is approximated using
finite differences.
To solve the linear system $A \mathbf{x}=\mathbf{b}$ with
$A=DG(\mathbf{u}^k)$, $\mathbf{b}=-G(\mathbf{u}^k)$ we used the Krylov subspace method \cite{Brown}.
The Krylov subspace of dimensionality $k$ is defined as
\begin{equation}
\mathcal{K}_k(A,\mathbf{v}_1)=\text{span} \{ \mathbf{v}_1,A \mathbf{v}_1,A^2 \mathbf{v}_1,\dots,A^{k-1} \mathbf{v}_1 \} \, .
\end{equation}
In the \textit{Generalized Minimum Residual} (GMRES) algorithm
the Krylov subspace is generated by $\mathbf{v}_1=\mathbf{r}_0/|\mathbf{r}_0|$ with 
$\mathbf{r}_0=A \mathbf{x}_0-\mathbf{b}$, and $\mathbf{x}_0$ an initial guess, see \cite{Brown}.
After $k$ iterations, the refined solution is given by
\begin{equation}
 \mathbf{x}_k=\mathbf{x}_0+V_k \mathbf{y}\, ,
\end{equation}
where the matrix $V_k=\left(\mathbf{v}_1,\dots, \mathbf{v}_k\right)$ has the base vectors 
of the Krylov subspace as its columns.
The vector $\mathbf{y}$ is chosen by minimizing  
the residuum
\begin{equation}\label{eq:A_min_app}
\lVert b-A \mathbf{x}_k \rVert_2 = \lVert r_0-A V_k \mathbf{y} \rVert_2 \stackrel{!}{=} \, \min \, , 
\end{equation}
where $\lVert . \rVert_2$ denotes the Euclidean norm.
For this procedure
an orthonormal basis of the Krylov subspace is generated with an Arnoldi process.
With the use of the similarity transformation
\begin{equation}
A V_k =V_{k+1} \tilde{H}_k \, ,
\end{equation}
where $\tilde{H}_k$ is an upper Hessenberg matrix, $\mathbf{v}_1=\mathbf{r}_0/|\mathbf{r}_0|$, and 
the orthogonality of $V_k$,
the optimality condition Eq.~(\ref{eq:A_min_app}) becomes
\begin{equation}
\lVert \tilde{H}_k \mathbf{y} -|\mathbf{b}|  \mathbf{e}_1 \rVert_2  \stackrel{!}{=} \, \min \, , 
\end{equation}
with $\mathbf{e}_1=\left(1,0,\dots,0\right)$ the first unit vector of dimension $k+1$. 
For a $\mathbf{y}$ that minimizes this norm the approximate solution
is given by $\mathbf{x}_k=\mathbf{x}_0+V_k \mathbf{y}$.
To improve the convergence of this iterative method 
preconditioning was used. A preconditioner $M$ is
multiplied to $A\mathbf{x}=\mathbf{b}$ such that $M^{-1}A$ is close to unity.
A preconditioner suitable for our model is the inverse of the
linear operator in Fourier space with a small shift $0<\epsilon \ll 1$
in order to avoid singularities i.e.
\begin{equation}
M=\left(\epsilon +\left(k^2-k_c^2 \right)^2+\frac{2}{\Delta t} \right)^{-1} \, .
\end{equation}
The convergence of Newton's method is only guaranteed from a
starting point close enough to a solution. In the integration scheme we use a
line search method to ensure also a global convergence \cite{Dennis}.
Newton's method Eq.~(\ref{eq:App_NewtonMethod}) is thus modified as
\begin{equation}
\mathbf{u}^{k+1}=\mathbf{u}^k+\lambda \Delta \mathbf{u}^k \, ,
\end{equation}
where the function
\begin{equation} 
f(\mathbf{u}^k)=\frac{1}{2}G(\mathbf{u}^k)G(\mathbf{u}^k) ,
\end{equation}
is iteratively minimized with respect to $\lambda$. \\
This integrator was implemented using the \textit{PetSc} library \cite{PetSc}.
As the dynamics converges towards an attractor an adaptive 
stepsize control is very efficient. The employed adaptive stepsize control
was implemented as described in \cite{Press}.
The described integration scheme has been generalized for an arbitrary number
of real or complex fields. The coupling terms are treated as additional nonlinearities in $N$. 
As a common intrinsic timescale we choose $T=tr_z$ with $r_z$ the bifurcation parameter
of the OP map.
Due to the spatial discretization not all points of the critical circle 
lie on the grid. Thus, the maximal growth rate on the discretized circle
is not exactly equal to $r$, the theoretical growth rate. 
In particular, some modes may be suppressed or even become unstable. 
Due to this we expect deviations from analytical solutions.
To minimize such deviations the size of the critical circle was chosen such that 
this disbalance between the active modes was minimized.
Periodic boundary conditions were applied to account for the translation invariance of the spatial pattern. 


\section*{Acknowledgments}
We thank Ghazaleh Afshar, Eberhard Bodenschatz, Theo Geisel, Min Huang, Wolfgang Keil, 
Michael Schnabel, Dmitry Tsigankov, and Juan Daniel Fl\'{o}rez Weidinger for discussions.




\listoffigures


\end{document}